\documentclass[12pt]{article}
\begin{document}
\input epsf
\def\be{\begin{equation}}
\def\bea{\begin{eqnarray}}
\def\ee{\end{equation}}
\def\eea{\end{eqnarray}}
\def\d{\partial}
\def\la{\lambda}
\def\eps{\epsilon}

\begin{titlepage}
\begin{center}
\hfill   hep-th/yymmnnn\\
\hfill EFI-06-06\\
\hfill NSF-KITP-06-29

\vskip 1cm

{\LARGE {\bf On gravitational description of Wilson lines
  }}

\vskip 1.5cm

{\large  Oleg Lunin }

\vskip 1cm

{\it Enrico Fermi Institute, University of Chicago,
Chicago, IL 60637
}

\vskip 3.5cm

\vspace{5mm}

\noindent

{\bf Abstract}

\end{center}

We study solutions of Type IIB supergravity, which describe the geometries dual 
to supersymmetric Wilson lines in ${\cal N}=4$ super--Yang--Mills. We show that 
the solutions are uniquely specified by one function which satisfies a Laplace 
equation in two dimensions. We show that if this function obeys a certain 
Dirichlet boundary condition, the corresponding geometry is regular, and we 
find a simple interpretation of this boundary condition in terms of D3 and D5 
branes which are dissolved in the geometry. While all our metrics have $AdS_5\times S^5$ 
asymptotics, they generically have nontrivial topologies, which can be uniquely specified 
by a set of non--contractible three-- and five--spheres.

\vskip 4.5cm

\end{titlepage}

\newpage

\section{Introduction.}
\renewcommand{\theequation}{1.\arabic{equation}}
\setcounter{equation}{0}

According to AdS/CFT correspondence \cite{mald,gkpw}, there exists a map 
between operators in 
${\cal N}=4$ 
SYM and states in string theory on $AdS_5\times S^5$. This map generically 
leads to a stringy state on the bulk side, however there is a nice class of 
BPS operators whose duals are well--described by the type IIB supergravity. 
Such states have been extensively analyzed in perturbation theory, where one 
considers linear excitations around $AdS_5\times S^5$. Computing correlation 
functions for these
perturbations, one finds a remarkable agreement with field theory results 
(see \cite{AdSrev} for the review). However as the conformal weight of operator
 in field theory becomes large, one should not expect that the linearized 
solution of supergravity gives a good approximation to the correct geometry, 
but one can still hope that for a wide class of semiclassical solutions, the 
stringy corrections are suppressed, and by solving nonlinear equations which 
follow from the lagrangian of SUGRA, one finds 
a good description of the bulk state. 

A concrete realization of this idea was given in \cite{llm}, where BPS 
geometries with 
$SO(4)\times SO(4)$ symmetry were constructed, and they were shown to have 
small curvature 
everywhere. Moreover, the properties of these geometries were in a perfect 
agreement with expectations coming from field theory, where the BPS states had an 
effective description in terms of free fermions \cite{jevicki,berenst}. The states 
analyzed 
in \cite{berenst} are parameterized in terms of their R charge $J$ and when it 
is small ($J\ll N$), the dual objects are perturbative gravitons\footnote{In 
fact, 
one has to consider an excitation of a coupled system which contains gravitons
 and five--form flux and 
but we will call these excitations "gravitons" to be short.}
When quantum number $J$ becomes comparable with number of three--branes $N$, 
the dual 
description is given in terms of curved D3 branes which are known as "giant 
gravitons" \cite{giant}, and the connection of these branes with field theory 
was discussed in \cite{hashHirItz}. Finally, as $J$ becomes much larger than $N$, the 
brane probe approximation breaks down, but for certain semiclassical 
states the geometric description can be trusted (the curvature invariants 
always remain finite), and corresponding metrics were constructed in 
\cite{llm}. It was shown that various charges computed on the geometric side were in 
a perfect agreement with corresponding quantities for the Fermi liquid. Moreover, a subsequent 
work \cite{mandal,maoz} showed that a semiclassical quantization of the geometries led to 
emergence of free fermions on the gravity side, thus providing a direct map between BPS 
states in field theory and the moduli space of the geometries.

\begin{figure}[htb]
\begin{center}
\epsfysize=2.2in \epsffile{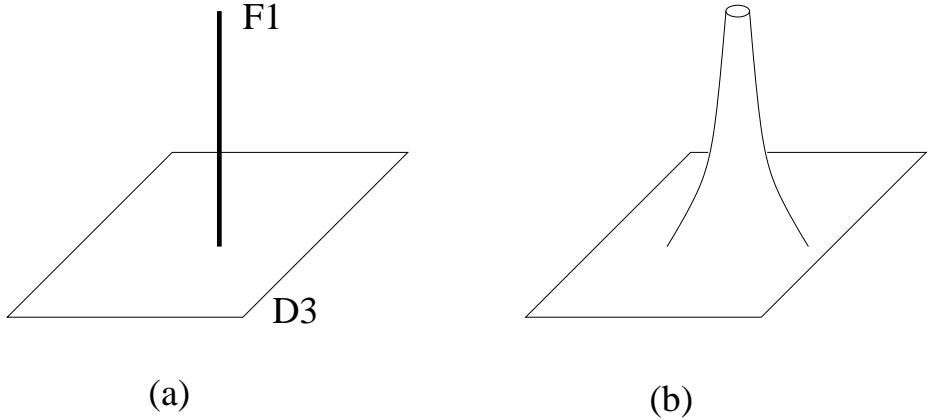}
\end{center}
\caption{Two different pictures for fundamental string ending on D3 brane: the naive 
configuration (a) and the description in terms of spike introduced in \cite{CalMald} (b). We 
will argue that the latter picture is responsible for existence of regular supergravity solution.
} \label{FigSpike}
\end{figure}

The goal of this paper is to develop a similar gravitational description for another class of BPS states. 
In the field theory such states are described by Wilson lines which break one--half of the 
supersymmetries. 
It is well--known that in AdS/CFT correspondence, to construct a dual bulk 
description of a Wilson line 
one considers an open fundamental string which ends on this line 
\cite{reyYee,maldLoop}. This picture should be 
true for the supersymmetric line as well, and this fact seems to imply that 
fundamental strings would 
always be present on the bulk side. However it is well--known that in a geometry 
produced by fundamental string, the dilaton and curvature invariants always diverge at the 
location of the string, so one concludes that Wilson lines cannot correspond to regular 
supergravity solutions in the bulk. Thus there 
seems to be a sharp contrast between these operators and the BPS states 
studied in \cite{berenst}: in the 
latter case the bulk description involved only D3 branes, and the resulting 
geometries were shown to be smooth \cite{llm}. In fact this difference in 
behavior is only an illusion, and as we review below, the brane configurations 
dual to Wilson lines should be viewed as D3 branes with fluxes rather than 
fundamental strings. This picture makes it plausible that in the geometric 
description, the dilaton stays finite and the metric remains regular, and our construction will 
show that this is indeed the case. Such "desingularization" is based on the effect discussed  
in \cite{CalMald}, where it was shown that a 
fundamental string ending 
on a D3 brane can be viewed as a curved D3 
brane which carries electric field 
(we illustrate it in figure \ref{FigSpike}). The relevance of this effect for the physics of 
Wilson lines was first proposed in \cite{reyYee}.
The argument of \cite{CalMald} was 
based on the dynamics of DBI 
action 
and to our knowledge its implications for supergravity solutions were 
never analyzed.  In this paper 
we show that in a certain setup (when supersymmetry is enhanced by going to 
the near horizon limit of D3 branes), the effect of \cite{CalMald} leads to 
regularization of supergravity solution, in particular the 
dilaton is bounded in the entire space.

In fact the supergravity analysis that we present here was partially performed in a nice paper 
\cite{yama}, which was an inspiration for the present work. However \cite{yama} gave only 
several necessary conditions for the geometry to be supersymmetric, and here we solve all 
supergravity equations. We show that all solutions with $AdS_5\times S^5$ asymptotics are 
parameterized by one harmonic function, and if this function obeys certain Neumann 
boundary conditions, the solution is guaranteed 
to be regular. Unfortunately to recover the metric, one still needs to solve some differential 
equations, but we prove that once the harmonic function is specified, this solution exists and 
it is unique. We also outline a perturbative procedure for constructing the solution.

However, before we discuss supergravity equations it might be useful to recall the 
description of BPS states in terms of the brane probes. This analysis is presented in section 
\ref{SectDBI}. In section \ref{SectSumma} we summarize the gravity solution (while the 
relevant algebra is presented in the appendices), and show that the geometry is regular. Section 
\ref{SectAdS} demonstrates that $AdS_5\times S^5$ solution can be easily recovered from the 
general formalism, and in section \ref{SectPert} we construct the perturbative series around this 
solution. The existence and uniqueness of this series proves that any harmonic function with 
correct boundary conditions unambiguously leads to the unique regular geometry. In section 
\ref{SectCont} we point out that once the complete system of equations is derived, it can be 
used to describe different brane configurations. In particular, in this paper we are interested 
in solutions 
with $AdS_2\times S^2\times S^4$ factors, but slight modifications of the system make it appropriate for describing geometries with $AdS_4\times S^2\times S^2$ factors. Such 
geometries are produced by backreaction of D5 branes with $AdS_4\times S^2$ worldvolumes.
It is curious that there exists another analytic continuation which maps our system back to itself, 
but in a different coordinate frame, and we discuss such continuation in section \ref{SectCont} 
as well. Finally in section \ref{SectGoBack} we discuss the topology of the solutions and we 
show that they admit some 3-- and 5--cycles, and by wrapping D3 or D5 branes of this cycles 
we recover the branes discussed in section \ref{SectDBI}.


\section{Wilson lines and brane probes.}
\renewcommand{\theequation}{2.\arabic{equation}}
\setcounter{equation}{0}


\label{SectDBI}

In this section we will summarize some known facts about the Wilson loops and
 brane configurations which are dual to them. Out goal is to construct the 
gravitational dual of supersymmetric time--like Wilson 
loops in ${\cal N}=4$ SYM. In field theory such operators are specified by a 
representation $R$ of the 
gauge group, and are given by the following expressions:
 \bea\label{WilsLine}
 W_R({\cal C})=\mbox{Tr}_R~P\exp\left(i\int_{\cal C}
ds(A_\mu{\dot x}^\mu+\phi_I{\dot y}^I)\right)
 \eea
Here the curve ${\cal C}$ is a straight line $x^0=t,{y}^m=n^m t$ and $n^m$ is 
a unit vector in $R^6$. 
The choice of this vector breaks $SO(6)$ R--symmetry down to $SO(5)$  which 
rotates the remaining five scalars. Before we introduced the Wilson loop, 
${\cal N}=4$ SYM had $SU(2,2)$ conformal symmetry, but the presence of the 
straight line breaks this symmetry down to $SU(1,1)\times SU(2)$ \cite{cardy}. 
Thus we expect that in the presence of Wilson loop, field theory has 
$SO(6)\times SU(1,1)\times SU(2)$ global symmetry, which implies that the 
gravity dual would contain $AdS_2$, $S^2$ and $S^4$ factors\footnote{The consequences of 
$AdS_2\times S^2$ symmetry for the field theory were recently studied in 
\cite{kapustin,yama,gomis}.}.

This is very reminiscent of the situation with BPS chiral primaries discussed 
in \cite{jevicki,berenst}: in 
that case field theory was defined on $R\times S^3$ and to construct a chiral 
primary one had to 
consider zero modes on the sphere. Moreover, a generic $1/2$ BPS state broke 
the R symmetry group 
down to $SO(4)$ (which is analogous to the $SO(6)\rightarrow SO(5)$ breaking 
for the Wilson lines), so 
on the bulk side the relevant symmetry was $SO(4)\times SO(4)\times U(1)$. The 
geometries for such BPS states were constructed in \cite{llm} (and the goal of 
this paper is to develop a similar picture for 
the states dual to Wilson lines (\ref{WilsLine})), but before that a great 
deal of information about the bulk 
states 
was extracted in the brane probe approximation. For the BPS states with 
$SO(4)\times SO(4)$ symmetry
the relevant branes were known as "giant gravitons" \cite{giant} and they were 
wrapping cycles either on 
$S^5$ (in that case the angular momentum was bounded from above:  $J\le N$) or 
on $AdS_5$. As we will show in this paper, there is a very close analogy 
between the geometries produced by giant gravitons
and the geometries which are dual to the Wilson lines (\ref{WilsLine}), so it 
is very natural to start with 
discussing the brane configurations which are dual to (\ref{WilsLine}).

We begin with the metric written in $AdS_2\times S^2\times S^4$ form:
\bea
ds^2&=&R^2\left(\cosh^2 \rho dH_2^2+d\rho^2+\sinh^2 \rho d\Omega_2^2+d\theta^2+
\sin^2\theta d\Omega_4^2\right)\\
F_5&=&4R^4\left(\cosh^2 \rho\sinh^2 \rho d\rho\wedge dH_2\wedge d\Omega_2+
dual\right)\\
R^4&=&4\pi Ng
\eea
According to the proposal of \cite{reyYee,fiol}, a dual description of the Wilson line
 is given by a D3 brane
with worldvolume $AdS_2\times S^2$, which is symmetric under $SO(5)$ rotations 
and has an electric field along 
its worldvolume.  In an analogy with giant gravitons, we assume that $\rho$ is 
fixed, this assumption is consistent with equations of motion. Then we look at 
the action for D3 brane:
\bea\label{D3DBI}
S_{D3}=-TR^4\int d^4\sigma\sqrt{(\cosh^4 \rho-E^2)\sinh^4 \rho}+4TR^4
\int d^4\sigma \int_0^\rho dv\cosh^2 v\sinh^2 v
\eea
Here $E$ is a value of an electric field on the worldvolume of the brane, to 
be consistent with symmetries and to have a closed form $F_{mn}$, this electric
 field must be constant. To simplify the expressions we defined a rescaled electric field 
$E=2\pi R^{-2}F_{01}$.
Extremizing the action with respect to $\rho$, we find an equation
\bea
-\frac{(1 - 4E^2 + 2\cosh 2\rho+\cosh 4\rho)\sinh 2\rho}{4\sqrt{-E^2+\cosh^4 
\rho}}+
4\cosh^2 \rho\sinh^2 \rho=0
\eea
This equation is always solved by $\rho=0$, and now we want to find another 
solution. Then the relevant equation becomes
\bea
4(\cosh^2 \rho\cosh 2\rho-E^2)=4\sqrt{-E^2+\cosh^4 \rho}\sinh 2\rho\nonumber
\eea
and it can easily be solved:
\bea\label{EnrgAsU}
E=\cosh \rho
\eea
We see that the value of the electric field cannot be smaller that one, and if 
$E=1$ we are back to the solution $\rho=0$. For $E>1$ we find two solutions: 
(\ref{EnrgAsU}) and $\rho=0$, they are 
counterparts of a giant graviton and a usual graviton in \cite{giant}. Just as 
in that case, one can look at a 
potential for $\rho$
and show that solution $\rho=0$ is unstable, and the correct expression is 
(\ref{EnrgAsU}). It may be more convenient to parameterize a brane by an electric 
displacement $\Pi$:
\bea\label{LeftGiant}
\Pi=\frac{\delta S}{\delta E}=\frac{TE}{\sqrt{\cosh^4 \rho-E^2}}\sinh^2 
\rho=T\sinh \rho
\eea
which goes to zero as the brane shrinks to zero size. Moreover, the electric 
displacement controls a coupling of the electric field $F_{01}$ with a bulk 
Kalb--Ramond field $B_{01}$ 
\cite{douglas,pawRey,camino}: to find this coupling in the linear order, one 
makes a substitution 
$2\pi F\rightarrow 2\pi F-B$ in (\ref{D3DBI}), then after performing an 
integration over $\Omega_2$, one finds 
a coupling
\bea\label{D3Bsource}
\delta S=\int d^4\sigma\frac{\delta S_{D3}}{\delta(2\pi F_{ab})}{\hat B}_{tx}=
\Omega_2
T_3 R^2\sinh\rho\int dt~dx~B_{tx}
\eea
Thus we see that, as expected, the three brane with electric flux sources a 
charge for fundamental string, and to extract the value of this charge, one 
has to solve equations of motion for $B$ with source (\ref{D3Bsource}), construct 
the relevant field strength $H=dB$, and integrate its dual over an appropriate 
manifold at infinity. Since, 
by construction, our string is uniformly smeared on $S^2$,  the manifold 
relevant for  the present case turns our to be $S^5$. Notice that going from 
the NS--NS three 
form $H$ to its dual and expressing the later in terms of unit sphere  $S^5$, 
ones introduces 
an extra factor of $R^2$, so the we find an expression for the number of 
fundamental strings:
\bea
\Omega_2 T_3 R^4\sinh\rho=\frac{n_f}{2\pi}:\qquad n_f=4N\sinh\rho
\eea
Since we are working in the units where $\alpha'=1$, we have the following expressions for the 
tension and the volume of the sphere:
\bea
T_3=\frac{1}{g(2\pi)^3},\quad \Omega_2=4\pi,\quad
\eea
In $AdS_5\times S^5$ there exists another brane which preserves 
$AdS_2\times S^2\times S^4$ 
symmetry: it is a D5 brane with worldvolume $AdS_2\times S^4$. In the Poincare 
patch, the solution for such probe brane was found in \cite{pawRey} and authors 
of \cite{gomis} explored the relation of this brane to Wilson lines (see also 
\cite{kumar} for an interesting discussion of non--BPS case). Let us 
see how such branes would look in global AdS. We again take an worldvolume field 
strength to be proportional to the volume of AdS space: $F=\frac{ER^2}{2\pi}d^2 H_2$, 
then the 
DBI action for D5 branes becomes
\bea
S_{D5}=-T_5 R^6\int d^6\sigma\sqrt{(\cosh^4 \rho-E^2)\sin^8 \theta}+4T_5 R^6
E\int d^6\sigma \int^\theta d\phi
\sin^4\phi
\eea
To have six--dimensional branes which preserve $S^2$ symmetry, one has to set 
$\rho=0$, then equation for $\theta$ leads to the relation
\bea
-4\sin^3\theta(\sqrt{1-E^2}\cos\theta-E\sin\theta)=0
\eea
Again, there is an unstable solution $\theta=0$, and the stable one
\bea\label{RightGiant}
&&\theta=\arctan\frac{\sqrt{1-E^2}}{E},\nonumber\\
&&\frac{\Pi}{T_5 R^6}=\sin^3\theta\cos\theta+
4\int d^6\sigma \int^\theta d\phi\sin^4\phi=\frac{3}{2}\left[\theta-
E\sqrt{1-E^2}\right]
\eea
As before, we can compute the number of fundamental strings generated by this solution.
 To this end we first find the relevant coupling to the $B$ field:
\bea
\delta S=\int d^6\sigma\frac{\delta S_{5}}{\delta(2\pi F_{ab})}{\hat B}_{tx}=
\frac{3}{2}\Omega_4
T_5 R^4\left[\theta-E\sqrt{1-E^2}\right]\int dt~dx~B_{tx}
\eea
Notice that in this case the strings are smeared over the four--sphere, so one 
needs to dualize $H_3$ in six dimensional space (the surface of integration is 
$S^3$), so no factors of $R$ are coming from the dualization. The number of 
strings is
\bea
n_f=2\pi\frac{3}{2}\Omega_4
T_5 R^4\left[\theta-E\sqrt{1-E^2}\right]=\frac{N}{\pi}\left[\theta-E\sqrt{1-E^2}
\right]
\eea
Here we used the expressions for the volume of the sphere and for the tension of the brane:
\bea
\Omega_4=\frac{8\pi^2}{3},\quad T_5=\frac{1}{(2\pi)^5g}
\eea

We observe that for a fixed value of displacement $\Pi$,  solutions of both 
(\ref{LeftGiant}) and (\ref{RightGiant}) exist, on the other hand, in terms of 
$E$ there 
seems to be a nice
 complementarity: if $E<1$ we only have the D5 solution, while for $E>1$ only D3 
solution is present. Of course, it is $\Pi$, not $E$ that is related to  physical 
observables, so the situation is quite analogous to the giant gravitons: for the 
same value of 
$\Pi$ we have a "giant" (D5 brane) and a "dual giant" (D3 brane). Notice that in 
the case of spherical branes, the angular momentum of the giant was bounded 
($J\le N$), while for dual giant it was not \cite{giant}, and we have a similar picture here: the 
value of $n_f$ in (\ref{RightGiant}) is bounded by $N$. 

The analogy between D5 brane and the giant graviton (and between D3 brane and the 
dual giant) is also supported by the field theory consideration of \cite{gomis}. 
First we recall that in the field theory, the chiral 
primaries of \cite{berenst} were constructed by taking various gauge invariant 
combination of a single $N\times N$ matrix $Z$. Such combinations can be 
classified in terms of representations of permutations group $S_N$ \cite{jevicki},
 in particular the giants correspond to antisymmetric representations and dual 
giants correspond to symmetric representations of this group \cite{vijay,jevicki}.
Recently a similar story emerged for the Wilson lines (\ref{WilsLine}): they are 
characterized by representations of the gauge group, and it was shown in 
\cite{gomis} that the symmetric representations correspond to D3 branes with 
fluxes, while antisymmetric representations correspond to D5 branes.

If we put many giant gravitons together, the brane probe approximation would break 
down and the geometry would no longer be $AdS_5\times S^5$. A generic configuration 
of giants would lead to a space with regions of large curvature, so one would need full 
string theory to describe such a space. However, there exist semiclassical configurations of 
giants which lead to regular geometries, and their metric would be well approximated by the 
solutions of type IIB supergravity. For the giant gravitons such solutions were constructed in 
\cite{llm}, and now we want to study similar semiclassical geometries for the D3 and D5 
branes which were discussed in this section. 

 
\section{Summary of supergravity solution.}
\renewcommand{\theequation}{3.\arabic{equation}}
\setcounter{equation}{0}


\label{SectSumma}

In this section we will outline the procedure for constructing the supergravity 
solutions, and the details of the computations are provided in the appendices. As 
we discussed in the previous section, the solution is expected to have $AdS_2$, 
$S^2$ and $S^4$ factors, so the metric and five--form are  given by
\bea\label{Metric0}
ds^2&=&e^{2A}dH_2^2+e^{2B}d\Omega_2^2+e^{2C}d\Omega_4^2+h_{ij} dx^i dx^j,\quad 
i,j=\{1,2\}\\
F_5&=&df_3\wedge d\Omega_4+*_{10}(df_3\wedge d\Omega_4)\nonumber
\eea
In the brane probe approximation, one set of the Wilson loops was described in 
terms of D3 branes with spikes of fundamental strings, and such fundamental 
strings produced NS--NS $B$ field along $AdS_2$ direction. Since we already have 
nonzero five--form, the equation
\bea
d*_{10}(e^\phi F_3)=gF_5\wedge H_3
\eea
implies that there is also a nontrivial RR potential $C^{(2)}_{\mu\nu}$ along the 
$S^2$ directions. Three form also sources the dilaton. In principle we could also 
have the RR form along $AdS_2$ and NS--NS form along $S^2$, but one can 
consistently set them to zero. To see this we notice that equations of motion of 
type IIB supergravity (but not the SUSY variations) are invariant under the change
 of sign of RR fields. Since we are looking for solutions with $S^4$ and $S^2$ 
factors, we also have a $Z_2$ symmetry which simultaneously inverts the 
orientation of these factors\footnote{Notice that if invert orientation of only 
one of the factors, then self--dual $F_5$ doesn't just flip sign, but it changes 
in a more complicated way.}. Combination of these two symmetries leaves $F_5$, 
$C^{(2)}_S$ and $B_{AdS}$ invariant, but it changes signs of 
$C^{(0)}$, $C^{(2)}_{AdS}$ and $B_{S}$. In the leading order, the fundamental 
string probe sources only 
$C^{(2)}_S$ and $B_{AdS}$, so this discrete symmetry guarantees that we can 
consistently set $C^{(0)}$, $C^{(2)}_{AdS}$ and $B_{S}$ to zero. To summarize, in 
addition to (\ref{Metric0}), we should excite the following fields:
\bea\label{3Form0}
H_3=df_1\wedge dH_2,\quad F_3=df_2\wedge d\Omega_2,\quad e^\phi
\eea
Notice that we started with string probe and used the symmetry argument to project
 out the unwanted components of three--forms, but we would have arrived to the 
same conclusion it we started from the D5--brane picture. 

We can now consider the supersymmetry variations of the ansatz (\ref{Metric0}), 
(\ref{3Form0}), and the details of this analysis are presented in the appendix. 
Notice that in the initial steps we are essentially repeating the arguments of 
\cite{yama}, although we are using more standard notation.
In the end we find that geometry can be expressed in terms of three fields 
$G,H,\phi$:
\bea
\label{MastEqnMetric}
&&ds^2=e^{2A}dH_2^2+e^{2B}d\Omega_2^2+e^{2C}d\Omega_4^2+
\frac{e^{-\phi}}{e^{2B}+e^{2C}}
(dx^2+dy^2)\\
\label{MastEqnFlux}
&&F_5=df_3\wedge d\Omega_4+*_{10}(df_3\wedge d\Omega_4),\quad
H_3=df_1\wedge dH_2,\quad F_3=df_2\wedge d\Omega_2\\
&&e^{2A}=ye^{H-\phi/2},\quad e^{2B}=ye^{G-\phi/2},\quad e^{2C}=ye^{-G-\phi/2},\quad 
F=\sqrt{e^{2A}-e^{2B}-e^{2C}}\nonumber\\
\label{MastEqnF1}
&&df_1=-\frac{2e^{2A+\phi/2}}{e^{2A}-e^{2B}}\left[e^A Fd\phi-e^{B+C}*d\phi
\right],\\
\label{MastEqnF2}
&&df_2=\frac{2e^{2B-\phi/2}}{e^{2A}-e^{2B}}\left[e^B Fd\phi-e^{A+C}
*d\phi\right]\\
\label{MastEqnF3}
&&e^B e^{-4C}*df_3=e^A d(A-\frac{\phi}{4})+\frac{1}{4}F e^{-\phi/2-2A}df_1
\eea
These fields satisfy two differential relations 
\bea\label{MastEqnD1}
&&d(H-G-2\phi)=-\frac{2}{y(e^{2B}+e^{2C})}(e^{2C}dy+ F e^{B+C-A}dx)\\
\label{MastEqnD2}
&&*d\arctan e^G+\frac{1}{2}d\log\frac{e^A-F}{e^A+F}-
\frac{1}{2}e^{-\phi/2-2A}df_1=0
\eea
along with integrability conditions coming from (\ref{MastEqnF1})--
(\ref{MastEqnF3})\footnote{Throughout this paper the star is used to denote a 
Hodge dual in a flat two dimensional space with coordinates 
$(x,y)$, and our convention is $*dy=dx$.}. Later we will also need an alternative form 
of equation (\ref{MastEqnF3}), which can be obtained by combining it with other equations 
in the system:
\bea\label{MastEqnF3Alt}
&&e^A e^{-4C}*df_3=e^B d(B+\frac{\phi}{4})+\frac{1}{4}F e^{\phi/2-2A}df_2
\eea

Notice that any solution of this system would lead 
to a supersymmetric geometry, in this sense we found the necessary and sufficient conditions 
for having a BPS solution. Unfortunately, 
we were not able to solve this system of equations. However it is clear that the 
entire problem can be 
reduced to one differential equation for one function: for example, equation 
(\ref{MastEqnD1}) allows us to 
express $G$ and $H$ in terms of 
derivatives of a function $\psi\equiv H-G-2\phi$. Then we also know $\phi$ as a 
function of $\psi$, and 
(\ref{MastEqnD2}) gives a closed differential equation for a single function 
$\psi$. Of 
course, this is a very inefficient way of solving the system, but it demonstrates 
that we essentially have one scalar degree of freedom. 

It is useful to introduce two more functions $\Psi_1$ and 
$\Psi_2$ by performing decomposition
\bea\label{DefPsi}
&&-\frac{1}{4}e^{-2A-\phi/2}df_1=\frac{e^{-\phi/2}}{2(e^{2A}-e^{2B})}\left[e^{A+\phi/2} 
Fd\phi-y*d\phi\right]
\equiv \frac{1}{2}(d\Psi_2+*d\Psi_1)
\eea
Of course there is an ambiguity in defining $\Psi_1$ and $\Psi_2$: one can add an 
arbitrary harmonic function to $\Psi_2$ and subtract the dual of this function 
from $\Psi_1$. We will use this ambiguity to impose the boundary condition
\bea\label{BCPsi2}
\Psi_1|_{y=0}=0,\quad \Psi_1|_{x^2+y^2\rightarrow\infty}=0
\eea
Let us rewrite equation (\ref{MastEqnD2}) in terms of $\Psi_1,\Psi_2$: 
\bea
*d\left[\arctan e^G+\Psi_1\right]+d\left[\frac{1}{2}\log\frac{e^A-F}{e^A+F}+
\Psi_2\right]=0
\eea
this implies that there exists a harmonic function $\Phi$ such that
\bea\label{MastHarmonic}
(\d_x^2+\d_y^2)\Phi=0:\qquad \arctan e^G+\Psi_1=\d_y\Phi,\quad
\frac{1}{2}\log\frac{e^A-F}{e^A+F}+\Psi_2=\d_x\Phi
\eea
As we argued before, the solution should be completely determined by one function,
 and we will specify the 
geometry by choosing $\Phi$\footnote{While it is true that geometry is specified 
in terms of one function, the choice of $\Phi$ does not fix a solution uniquely: 
this function turns out to be invariant under the constant shifts of the dilaton 
(while keeping $G,H,x,y$ fixed). To deal with this ambiguity we notice that 
if the dilaton is rescaled as  $e^{\phi'}=ge^{\phi}$, then 
equations (\ref{MastEqnF1})--(\ref{MastEqnD2}) remain invariant provided that
$$x'=gx.\quad y'=gy, \quad f_1'=g^{1/2}f_1,\quad f_2'=g^{-1/2}f_2,\quad f_3'=f_3$$
In this paper we fix the value of $e^\phi$ at infinity to be one, then $\Phi$ 
indeed corresponds to a unique solution. The string coupling constant can be 
recovered by making the rescaling written above.}. 
While we do not have explicit expressions for the metric components in terms of 
 $\Phi$, we still can determine the correct boundary conditions for this function,
 and in the next section we will outline the perturbative procedure for 
constructing geometry for any $\Phi$.

Having formulated the local differential equations, we will now discuss the 
relevant boundary conditions. 
Since we are looking at solutions which are dual to BPS states in field theory, 
we expect the geometries to be regular. We recall that another class of BPS 
geometries was discussed in \cite{llm} where it was shown that locally the metric 
can be expressed in terms of a single harmonic function. Then regularity led to 
particular boundary conditions for this function. In the present case, we already 
saw that the solution can also be specified in terms of one harmonic function, and
 now we will show that regularity imposes very simple boundary conditions for 
$\Phi$.
 
\begin{figure}
\begin{center}
\epsfxsize=4.5in \epsffile{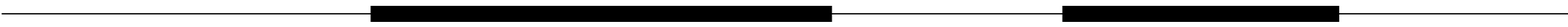}
\end{center}
\caption{A pictorial representation of the boundary conditions (\ref{PhiBC}) on $y=0$ 
line: the dark region corresponds to shrinking $S^2$ and the light regions correspond 
to contracting $S^4$. Since we are looking for solutions with $AdS_5\times S^5$ 
asymptotics, the dark segments are contained in a finite region of the line.
} \label{FigDivLine}
\end{figure}

The geometry (\ref{MastEqnMetric}) has two spheres and the product of their 
radii is equal to 
$ye^{-\phi}$,
while the ratio of the radii is $e^G$. So if one of the spheres goes to zero size,
  then $G$ approaches either positive or negative infinity, while $y$ goes to 
zero. Since the ambiguity in $\Psi_1$ was fixed by 
(\ref{BCPsi2}), equation (\ref{MastHarmonic}) leads to two kinds of boundary 
conditions:
\bea\label{PhiBC}
&&\d_y\Phi |_{y=0}=\frac{\pi}{2}:\quad S^4\ \mbox{shrinks}\nonumber\\
&&\d_y\Phi |_{y=0}=0:\quad S^2\ \mbox{shrinks}
\eea
Thus the line $y=0$ is divided into the set of regions where normal derivative of 
$\Phi$ has a certain value. This is analogous to the picture of \cite{llm} where 
there was a harmonic function with two kinds of Dirichlet boundary conditions in 
the plane. Pictorially the line $y=0$ is shown in figure \ref{FigDivLine}, where dark regions 
correspond to shrinking $S^2$ and light regions correspond to shrinking $S^4$. 
Let us assume that the dark regions are given by $x_{2m-1}<x<x_{2m}$, then we can 
find the complete solution of the Laplace equation:
\bea\label{MastSolut}
\Phi&=&\frac{\pi y}{2}-\frac{1}{4}\sum\int_{x_{2m-1}}^{x_{2m}}d\xi\log[(x-\xi)^2+
y^2]\nonumber\\
&=&\frac{\pi y}{2}+\frac{1}{4}\sum\left[-2(x-\xi)+2y\arctan\frac{x-\xi}{y}+(x-\xi)
\log[(x-\xi)^2+y^2]
\right]_{x_{2m-1}}^{x_{2m}}\nonumber\\
\d_y\Phi&=&\frac{\pi}{2}+\frac{1}{2}\sum
\left(\arctan\frac{x - x_{2m}}{y} -\arctan\frac{x - x_{2m-1}}{y}\right)\\
\d_x\Phi&=&\frac{1}{4}\sum \log\frac{(x-x_{2m})^2+y^2}{(x-x_{2m-1})^2+y^2}
\nonumber
\eea


Notice that we can add an arbitrary function of $x$ to $\Phi$, and we fixed this 
freedom by requiring that the derivative $\d_x\Phi$ goes to zero as $y$ goes to 
infinity. Here we also assumed that the dark segments are concentrated in a finite 
region of $y=0$ line. This is required for the solution to be asymptotically 
$AdS_5\times S^5$, and in this paper we will only be interested in such solutions.

Let us now show that any harmonic function $\Phi$ which has boundary conditions 
(\ref{PhiBC}) in the various regions of $y=0$ line, leads to a regular geometry. 
It is clear that once we stay away from $y=0$ and infinity of $(x,y)$ plane, all 
coefficients in the metric remain finite, so the geometry is regular.  So we only 
need to analyze the behavior of the metric near $y=0$ and at infinity. As we will see in the next section, the metric based on harmonic function (\ref{MastSolut}) approaches 
$AdS_5\times S^5$ at infinity, so the space is clearly regular there. We will now analyze the 
points on the $y=0$ line.

We begin with vicinity of the points in the dark region (i.e. we look near 
$(x,y)=(x_0,0)$, where $x_{2m-1}<x_0<x_{2m}$).
Then we can expand $\Phi$ and equations (\ref{MastHarmonic}), (\ref{MastEqnD1}) 
become  
\bea\label{ApproxMast}
&&e^G+\Psi_1=yq_1(x),\quad \frac{1}{2}\log\frac{e^A-F}{e^A+F}+\Psi_2=q_2(x)
\nonumber\\
&&d(H-G-2\phi)=-\frac{2dy}{y}-\frac{2Fe^{-A}}{ye^{-G}}dx
\eea
The first equation shows that generically $e^G\sim y$, $\Psi_1\sim y$, then 
integrability of the 
last equation implies that in the leading order\footnote{To arrive at this 
conclusion one should also recall that
$e^H\ge e^G+e^{-G}\sim y^{-1}$}
\bea
Fe^{-A}={\tilde q}_3(x) ye^{-G}=q_3(x)
\eea
Substituting this into the second equation in (\ref{MastHarmonic}), we find that 
in the 
leading order, $\Psi_2=\Psi_2(x)$.
Using all this information, we can write the leading contribution to the 
equation for the dilaton:
\bea
-\frac{1}{2}e^{-2A-\phi/2}df_1=q_3(x)d\phi-e^{-H}*d\phi=d\Psi_2+*d\Psi_1
\equiv dp_1(x)+O(y)
\eea 
This leads us to the important conclusion that neither dilaton nor $f_1$ diverges 
as we approach $y=0$. This behavior should be contrasted with gravitational 
solution for fundamental string which has a divergent dilaton. So the gravity 
solution confirms the picture which we discussed in the previous section: rather than having 
the "naked" sources of fundamentals strings, the geometry is described by regular D3 
branes and the string charge is mimicked by the fluxes on the brane. 

Looking at the equation for $f_2$, we wind that
\bea
df_2\sim -2e^{2B-\phi/2-A+C}*d\phi\sim y *d\phi
\eea
Thus the potential $f_2$ scales like the volume of $S^2$ which is necessary for 
having a regular solution. Finally we analyze the metric. Since dilaton remains 
finite, we can find the leading expressions for the warp factors:
\bea
e^{2B}=a_1(x)y,\quad e^{2C}=a_2(x),\quad e^{2A}=a_3(x)
\eea
In other words, the radii of $S^4$ and $AdS_2$ remain finite
and the metric in $(S^2,x,y)$ sector remains regular:
\bea\label{GrowFlat}
e^{-\phi/2}ye^G d\Omega_2^2+\frac{e^{-\phi/2}}{y(e^G+e^{-G})}(dx^2+dy^2)\sim
e^{-\phi/2}q(x)\left[dy^2+y^2 d\Omega_2^2+dx^2 \right]
\eea

The vicinity of the points where $\d_y\Phi=-\frac{\pi}{2}$ can be analyzed in the 
analogous fashion. The counterparts of the equations (\ref{ApproxMast}) are
\bea
&&-e^{-G}+\Psi_1=yq_1(x),\quad \frac{1}{2}\log\frac{e^A-F}{e^A+F}+\Psi_2=q_2(x)
\nonumber\\
&&d(H-G-2\phi)=-\frac{2e^{-2G}dy}{y}-\frac{2Fe^{-A}}{ye^{G}}dx
\eea
With trivial modification of the arguments presented above, we find 
$e^{-G}\sim y$, while the dilaton, fluxes and the $AdS$ warp factor remain finite.
 To show the regularity of the metric we then only need to analyze the $(S^4,x,y)$
 sector of the geometry, and these coordinates combine to give a locally flat six 
dimensional space similar to (\ref{GrowFlat}). 
Finally, at the points where both spheres shrink to zero size, the geometry is 
also regular, and the simplest way to see this is to "zoom in" on such point by rescaling coordinates. Doing this one concludes that $(S^2,S^4,x,y)$ combine to form a patch of flat 
eight--dimensional space, which proves the regularity of the geometry.

To summarize, we proved that the problem of finding BPS supergravity solution is 
reduced to solving equations (\ref{MastEqnF1})--(\ref{MastEqnD2}). We also 
demonstrated that the solutions 
can be specified in terms of one function, and the most convenient way to 
parameterize the solution is to introduce a harmonic function $\Phi$ by 
(\ref{MastHarmonic}). To describe physical situation (e.g. to avoid imaginary 
values of 
$e^G$) this function should satisfy a simple Neumann boundary conditions 
(\ref{PhiBC}) on a line $y=0$.  We showed that any harmonic function which obeys 
these conditions leads to a regular geometry, in particular, dilaton always 
remains finite. Unfortunately,
in order to translate information from the harmonic function to the geometry one still
 needs to solve differential equations. In the section \ref{SectPert} we will present an 
algorithm which allows one to start from any function $\Phi$ and construct gravity
 solution as a perturbative expansion in the value of dilaton. Since dilaton goes 
to zero at infinity (asymptotically the space is $AdS_5\times S^5$) and never 
diverges, we expect that such perturbation theory should give convergent series 
rather than asymptotic expansion. We present a perturbative procedure for two 
reasons. First, it is interesting to look at the leading order correction to AdS 
space. But more importantly, our argument that the solution in completely 
determined in terms of $\Phi$ was somewhat formal, and perturbative expansion 
proves this statement by construction. But before we construct the perturbative 
series, it is useful to recover $AdS_5\times S^5$ space itself.


\section{Example: $AdS_5\times S^5$.}
\renewcommand{\theequation}{4.\arabic{equation}}
\setcounter{equation}{0}


\label{SectAdS}

Once we found the equations which describe all BPS geometries, it is interesting 
to see how  
$AdS_5\times S^5$ fits into the general story. As we mentioned in the previous 
section, the solution should be completely specified in terms of one function, 
and since $AdS_5\times S^5$ has vanishing dilaton\footnote{See footnote 5 for 
the discussion of constant dilaton}, we expect that this would be the only 
solution possessing such property. It is instructive to show that this is indeed 
the case. First we observe that 
there is an alternative form of equations (\ref{MastEqnD1}), (\ref{MastEqnD2}) (see 
appendix \ref{AppSUGRA} for details):
\bea\label{AltEqn1}
&&\frac{1}{2}d[e^{2A}-e^{2B}]-\frac{1}{4}(e^{2A}+e^{2B})d\phi+\frac{1}{2}Fe^A 
e^{-\phi/2-2A}df_1+\frac{F^2}{2} d\phi=0
\\
\label{AltEqn2}
&&e^{B+C} *d(C-\frac{\phi}{4})-\frac{1}{4}e^{2B} e^{-\phi/2-2A}df_1-
F e^A d(A-\frac{\phi}{4})-\frac{1}{4}F^2 e^{-\phi/2-2A}df_1=0
\eea
If dilaton is equal to zero, then according to (\ref{MastEqnF1}), $f_1$ vanishes 
as well,
 and equation (\ref{AltEqn1}) implies that
\bea
e^{2A}-e^{2B}=L^2,\quad e^{2C}=L^2-F^2<L^2
\eea
with constant $L$. These equations guarantee that we can parameterize warp factors
 in terms of two scalar functions, which we call $\rho$ and $\theta$:
\bea
e^{2A}=L^2\cosh^2\rho,\quad e^{2B}=L^2\sinh^2\rho,\quad e^{2C}=L^2\sin^2\theta
\eea
At this point we know that $y=L^2\sinh\rho\sin\theta$, however $x$ is still 
undetermined. To find an expression for it, we need to use the duality relation 
$dx=*dy$. In particular, we need the expression for $*d\rho$ and 
$*d\theta$, and those are provided by the equation (\ref{AltEqn2}):
\bea
\frac{1}{\sqrt{1-e^{2C}}} *de^C=e^{-B}de^A:\qquad *d\theta=d\rho
\eea
This information allows us to find the expression for $x$, and plugging it into 
the general expression for the metric (\ref{MastEqnMetric}), we recover 
$AdS_5\times S^5$ 
space\footnote{The change of variables from $(\rho,\theta)$ to $(x,y)$ was found 
before in \cite{yama} by starting from $AdS_5\times S^5$ solution and combining 
the warp factors to produce $y$ and $x$ coordinates. In contrast to this approach 
of matching parameters, we derive this solution, and more importantly, we find a 
connection to the harmonic function.}:
\bea
ds^2&=&L^2(\cosh^2\rho ds_{AdS}^2+\sinh^2 \rho ds_2^2+\sin^2\theta ds_4^2+d\rho^2+
d\theta^2)\\
\label{xyAdS}
&&x=L^2\cosh\rho \cos\theta,\quad y=L^2\sin\rho \sin\theta
\eea 
Once we know the warp factors as functions of $x$ and $y$, we can use equations 
(\ref{MastHarmonic}) to recover the harmonic function $\Phi$ which corresponds 
to 
$AdS_5\times S^5$ solution. The result turns out to be in the form 
(\ref{MastSolut}) with 
only one dark region with $x_2=-x_1=L^2$:
\bea\label{PhiForAdS}
\Phi=\frac{\pi y}{2}+\frac{1}{4}\left[2y\arctan\frac{x-\xi}{y}+(x-\xi)\log[(x-\xi)^2+y^2]
\right]_{\xi=-L^2}^{\xi=L^2}
\eea
This is analogous to the way in which 
$AdS_5\times S^5$ arose as a "bubbling solution" of \cite{llm}, where it 
corresponded to a harmonic function with sources in a circular dark region. 

Starting from $AdS_5\times S^5$ space we can recover the flat space in three different ways, and all of them would correspond to singular limits since one is changing the asymptotics. The first two ways are similar to the recovery of flat space from the bubbling 
solutions of \cite{llm}: we decompactify one of the spheres by taking some point on the $y=0$ line and rescaling coordinates to zoom in on this point. For example, if we look near the point 
in the dark strip, then it is metric of $S^4$ that has to be rescaled by infinite factor (and metric of 
$AdS_2$ is rescaled as well), so we end up with space where directions
along $S^4$ and $AdS_2$ became flat, while sphere $S^2$ combines with $y$ direction to give
an $R^3$. Near the point in the light region, $S^2$ and $S^4$ exchange roles. The third way to 
obtain flat space is to look at the vicinity of the point where dark region merges with light one, 
and in \cite{llm} such points led to pp wave metrics. However, in the present case, such nontrivial limit does not exist, and the only way to obtain a regular geometry is to go all the way to flat space by decompactifying $AdS_2$ and combining $(S^2,S^4,x,y)$ into $R^8$.


\section{Perturbative solution.}
\renewcommand{\theequation}{5.\arabic{equation}}
\setcounter{equation}{0}

\label{SectPert}

While we were not able to solve the differential equations in the complete 
generality, it might be interesting to look at special cases where they allow 
some analytic treatment. In the previous section we considered a particular 
solution corresponding to $AdS_5\times S^5$ space and it might be interesting to 
develop a perturbation theory around this solution. While such perturbative 
solution is interesting by itself (for example, its properties can be compared 
with CFT computations for Wilson loops), in our case it would play an important 
role in demonstrating that the gravity solution exists for any function $\Phi$. 
In section \ref{SectSumma} we gave a heuristic argument that all solutions have 
to be 
parameterized in terms of a single function and then we claimed that $\Phi$ can 
be viewed as such function. Since we are interested in solutions that asymptote 
to  $AdS_5\times S^5$, function $\Phi$ would be approaching (\ref{PhiForAdS}) as 
we go to infinity of 
$(x,y)$ plane, in particular the dilaton would approach zero. Then starting from 
large values of $(x,y)$ one can start doing perturbation theory in the value of 
$\phi$, and as we show in this section, every harmonic function $\Phi$ defines a 
unique perturbative series. Alternatively, this series can be viewed as an 
expansion in powers of $1/\sqrt{x^2+y^2}$ and certainly it has a nonvanishing 
radius of convergence. While we do not show that the series converges in the 
entire plane, we expect the metric components and the fluxes to be analytic, so once 
we show that there is a unique solution in the asymptotic region we expect that it can be unambiguously continued to the upper--half plane, and the resulting solution is 
guaranteed to be regular by the arguments of section \ref{SectSumma}.

Let us begin with equations (\ref{MastEqnF1})--(\ref{MastHarmonic}) and look 
at them at large 
values of radial coordinate 
in $(x,y)$ plane.
If space asymptotes to $AdS_5\times S^5$ (i.e. if all points $x_m$ in 
(\ref{MastSolut}) 
are bounded as 
$|x_m|<x_0$), then at large distances function $\Phi$ approaches (\ref{PhiForAdS})
 and the deviation would lead to a small correction to $AdS_5\times S^5$. Let us 
introduce a small parameter $\eps$ and write all functions as expansion in its 
powers:
\bea\label{AsympExpans}
&&G=G_{(0)}+\sum_1^\infty \eps^m g_{(m)},\quad  H=H_0+\sum_1^\infty \eps^m 
h_{(m)},\quad
\phi=\sum_1^\infty \eps^m \phi_{(m)},\\
&&\Phi= \Phi_{AdS}+\eps(\Phi-\Phi_{AdS})\equiv \Phi_{(0)}+\eps\Phi_{(1)}\nonumber
\eea
Here quantities with subscript zero correspond to $AdS_5\times S^5$, and by 
definition, the series for $\Phi$ has only one term. Let us look at equation 
(\ref{DefPsi}) in the $m$-th order:
\bea\label{MthOrder}
d\Psi^{(m)}_2+*d\Psi^{(m)}_1=xd\phi_{(m)}-y*d\phi_{(m)}+\dots=
d(x\phi_{(m)})-*d(y\phi_{(m)})+\dots
\eea
Here dots represent the terms containing expressions with orders between one and 
$m-1$. The terms with 
$\phi_{(m)}$ turned out to be remarkably simple, in particular due to the relation
\bea
dx=*dy
\eea
we were able to decompose it into exact and co--exact forms.  Suppose we solved 
the equations for all orders up to $m-1$--th, then in (\ref{MthOrder}) we know all
 terms represented by dots explicitly, and we can decompose them into exact and 
co--exact forms. This implies that\footnote{Notice that $y\phi_{(m)}$ vanishes at 
$y=0$, this means that to be consistent with (\ref{BCPsi2}), we should work in a 
gauge
 where ${\tilde \Psi}^{(m)}$ vanishes there as well. This is especially important 
for the first order since it allows us to choose 
${\tilde \Psi}^{(1)}_1=0$.}
\bea\label{MthOrderOne}
\Psi^{(m)}_2=x\phi_{(m)}+{\tilde \Psi}^{(m)}_2,\quad \Psi^{(m)}_1=-y\phi_{(m)}+
{\tilde \Psi}^{(m)}_1
\eea
These expressions should be substituted into the $m$--th order of equations 
(\ref{MastEqnF1})--(\ref{MastHarmonic}), but in addition 
we should expand the terms with $G$ and $H$ in powers of epsilon. Since terms in 
orders less than $m$ are known at this point, we can move them to the right hand 
side, and the contribution of the $m$--th order is evaluated in the appendix 
\ref{AppPert}. In the end we find:
\bea\label{MthOrderEqn}
&&\frac{g^{(m)}}{s^2+sh^2}-\phi^{(m)}=\frac{1}{y}\d_y\Phi^{(m)},\nonumber\\ 
&&h^{(m)}-g^{(m)}-2\phi^{(m)}+4s^2\phi^{(m)}=-\left\{\frac{2s^2}{y}\d_y\Phi^{(m)}+
\frac{2c^2}{x}\d_x\Phi^{(m)}\right\}
\eea
Here $\Phi^{(m)}$ contains the contributions from lower orders (the only exception
 is $\Phi^{(1)}$ which was defined in (\ref{AsympExpans})), and thus it is known 
explicitly. We also introduced a shorthand notation:
\bea
sh=\sinh\rho,\quad ch=\cosh\rho,\quad s=\sin\theta,\quad c=\cos\theta
\eea
and expressions for these quantities can be obtained by inverting (\ref{xyAdS}). 
Equations (\ref{MthOrderEqn}) allow us to express $m$--th order solution in terms 
of one unknown function (for example, $g^{(m)}$), and to determine this function 
we need more equations. In particular, we can take the $y$ component of equation 
(\ref{MastEqnD1}):
\bea\label{YeqnMord}
\d_y(h^{(m)}-g^{(m)}-2\phi^{(m)})=\frac{4y\phi^{(m)}}{s^2+sh^2}+
\frac{4\d_y\Phi^{(m)}}{s^2+sh^2}+
\Psi_y^{(m)}
\eea
where $\Psi_y^{(m)}$ contains contributions from lower orders. Using the relations
\bea
\d_y\theta=\frac{sh~c}{sh^2+s^2},\quad \frac{y}{s^2+sh^2}=-\d_y\log c,\quad 
\frac{1}{s^2}d(s^2-\log c)=\frac{1}{sc}(2c^2+1)d\theta=d\log\frac{s^3}{c}\nonumber
\eea
we arrive at the final equation
\bea\label{FiEqnMord}
&&\frac{c}{s}\d_y\left(\frac{s^3}{c}\frac{g^{(m)}}{2(s^2+sh^2)}\right)=
\frac{1}{4}\d_y\left\{\frac{s^2}{y}\d_y\Phi^{(m)}-\frac{c^2}{x}\d_x\Phi^{(m)}
\right\}-\frac{1}{8}\Psi_y^{(m)}
\eea
This completes the proof that starting from any harmonic function $\Phi$ which has
 the same asymptotics as 
$\Phi_{AdS}$, we can construct a unique perturbative series around 
$AdS_5\times S^5$ solution, and this series approximates the solution 
corresponding to $\Phi$ at large distances. We also expect that the series for the
 dilaton converges everywhere and yields the solution corresponding to $\Phi$, 
while series for $G$ and $H$ should converge away from the line $y=0$. 


\section{Analytic continuations.}
\renewcommand{\theequation}{6.\arabic{equation}}
\setcounter{equation}{0}

\label{SectCont}

The goal of this paper is an exploration of supersymmetric geometries with 
$AdS_2\times S^2\times S^4$ symmetries, however once we derived the main result 
(\ref{MastEqnMetric})--(\ref{MastEqnD2}), it can be used for describing some other 
solutions as well. In 
particular, we arrived at 
$AdS_2\times S^2\times S^4$ factors by analyzing the field theory configurations 
and recalling the observations of \cite{cardy} that a one--dimensional Wilson line 
breaks the conformal group in four dimensions down to $SO(2,2)\times SU(2)$. 
Similarly, studying domain walls in field theory, one is naturally led to 
$AdS_3\times S^1$ split of the four dimensional space. It turns out that $N=4$ 
SYM on this space and on $R\times S^3$ has similar structure of BPS states: all 
of them preserve $SO(4)$ R--symmetry group. In fact, the geometries dual to BPS 
states on $AdS_3\times S^1$ were constructed in \cite{llm} by making a certain 
analytic continuation of metrics dual to states in $R\times S^3$. In the present 
context, it is very natural to ask whether a similar analytic continuation leads to 
any interesting statements. 

By an analogy with analytic continuation of \cite{llm}, one may think about 
exchanging $AdS_2$ and $S^2$ factors in the solution. This can be accomplished
 by the following replacements:
\bea\label{AnalSS}
ds^2_{AdS}\leftrightarrow -ds_S^2,\quad x\rightarrow ix',\quad y\rightarrow iy',
\quad
G\rightarrow G'+\frac{\pi i}{2},\quad
H\rightarrow H'+\frac{\pi i}{2},
\eea
Substituting this into the equations of motion, we find that after rescaling the 
fluxes $f_1$, $f_2$ and redefining $F$ as
\bea\label{AnalContFlux}
f_1\rightarrow if_1',\quad f_2\rightarrow if_2',\quad F\rightarrow F'\equiv
\sqrt{e^{2B'}-e^{2A'}-e^{2C'}}
\eea
we arrive at the system for the real primed variables:
\bea\label{AnlCntD1}
&&d(H'-G'-2\phi)=-\frac{2}{y'(-e^{2B'}+e^{2C'})}(e^{2C'}dy'+ 
F' e^{B'+C'-A'}dx)\\
\label{AnlCntF1}
&&\frac{1}{2}e^{-2A'-\phi/2}df'_1=\frac{1}{-e^{2A'}+e^{2B'}}\left[e^{A'} F'd\phi-
e^{B'+C'}*d\phi\right],\\
\label{AnlCntF2}
&&-\frac{1}{2}e^{-2B'+\phi/2}df'_2=\frac{1}{-e^{2A'}+e^{2B'}}\left[
e^{B'} F'd\phi-e^{A'+C'}*d\phi\right]\\
\label{AnalContF3}
&&e^{B'} e^{-4C'}*df_3=e^{A'} d(A'-\frac{\phi}{4})-\frac{1}{4}F' e^{-\phi/2-2A'}df'_1\\
&&*d~\mbox{arctanh}~ e^{G'}-\frac{i}{2}d\log\frac{ie^{A'}-F'}{ie^{A'}+F'}+
\frac{1}{2}e^{-\phi/2-2A'}df'_1=0\nonumber
\eea
Notice that we have not simplified the last equation to make its origin more 
transparent, but once simplification is done, the factors of $i$ disappear from 
that equation:
\bea\label{AnalContTheEqn}
&&\frac{1}{2} *d\log\frac{e^{G'}-1}{e^{G'}+1}-d\arctan\frac{e^{A'}}{F'}+
\frac{1}{2}e^{-\phi/2-2A'}df'_1=0
\eea
One can worry that the system of equations written above does not make sense in 
type IIB supergravity since (\ref{AnalContFlux}) seems to suggest that real values
 of $f_1'$ lead to imaginary fluxes in the original solution and vice versa. 
However this is not the case. To see this we recall the expression for the 
complex three--form $G_3$ in the original variables 
\bea\label{G3SPredual}
G_3=e^{-\phi/2}H_3+ie^{\phi/2}F_3=e^{-\phi/2}df_1\wedge dH_2+
ie^{\phi/2}df_2\wedge d\Omega_2
\eea
In terms of primed variables this expression becomes
\bea\label{G3Sdual}
G_3=
ie^{\phi/2}df'_2\wedge dH'_2-e^{-\phi/2}df'_1\wedge d\Omega'_2
\eea
Here we used the following conventions for continuing the volume factors (
$ds_S^2\leftrightarrow -ds_H^2$ did not specify the continuation uniquely):
\bea
d\Omega_2=-idH_2',\quad dH_2=id\Omega'_2
\eea
We see, that in the new variables there is a NS--NS magnetic and RR electric
fields, i.e. we ended up with configuration of NS5 and D1 branes which is S--dual to
the one we started with. Comparing (\ref{G3SPredual}) and 
(\ref{G3Sdual}), we conclude that the after analytic continuation, the dilaton is 
$\phi'=-\phi$, which is consistent with S duality.

If double analytic continuation leads us to the equivalent system, one may wonder
 why the equations 
(\ref{AnlCntD1})--(\ref{AnalContTheEqn}) are different from the original system 
(\ref{MastEqnF1})--(\ref{MastEqnD2}). To be precise, there are some similarities: in particular 
it we write 
the equations for the fluxes (\ref{MastEqnF1}), (\ref{MastEqnF2}) in terms of 
$\phi'=-\phi$, then they go into (\ref{AnlCntF1}) and (\ref{AnlCntF2}) after replacements
\bea\label{SdualityRule}
A' \rightarrow B,\quad B' \rightarrow A,\quad \phi' \rightarrow \phi,\quad 
f'_1\rightarrow - f_2,
\quad f'_2\rightarrow - f_1
\eea
Also equation (\ref{AnalContF3}) goes into (\ref{MastEqnF3Alt}) under the same 
replacement. However the two remaining equations (\ref{AnlCntD1}), 
(\ref{AnalContTheEqn}) look very 
much different from their counterparts (\ref{MastEqnD1}), 
(\ref{MastEqnD2}). This difference is an artifact of our coordinate choice:  in the 
original frame we defined 
$y=e^{B+C+\phi/2}$, but tracing the fate of $y'$ under the map (\ref{SdualityRule}),
we find\footnote{Notice that in the Appendix \ref{AppSUGRA} we used a combination of
(\ref{App2proj1}) to argue that $y$ was a convenient coordinate. Alternatively, we could 
add (\ref{App1eqnA}) to (\ref{App1eqnC}), this would naturally lead to $y'$.}
that it goes into $e^{A+C+\phi/2}$, i.e. we have a description of the same system, 
but in a different coordinate frame. While there is nothing wrong in defining a coordinate 
$y'=e^{B'+C'+\phi/2}$, if 
we do so the points where $S^2$ shrinks to zero size would be somewhere in the 
middle of $(x',y')$ plane and now we will discuss the constraints which come from 
the regularity conditions at these points. 

\begin{figure}
\begin{tabular}{cc}
(a)&
\epsfxsize=4.5in \epsffile{DVDLine.eps}
\end{tabular}
\begin{center}
\epsfxsize=4.5in \epsffile{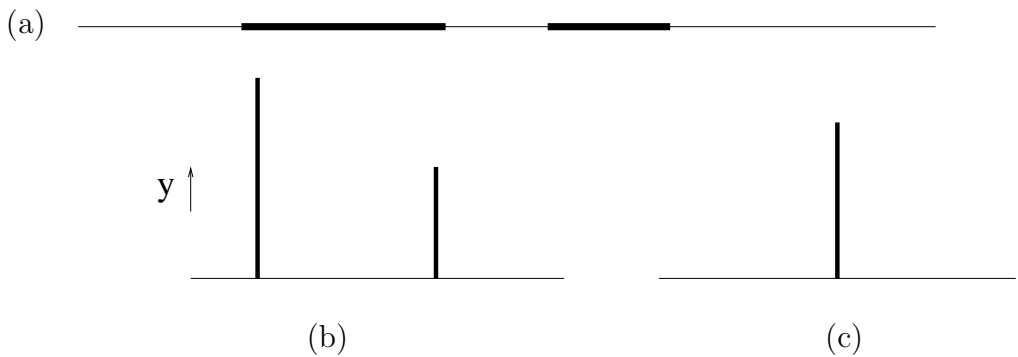}
\end{center}
\begin{tabular}{ccc}
\hskip 4cm (b)&& \hskip 5.5cm (c)
\end{tabular}
\caption{A pictorial representation of the boundary conditions (\ref{PhiBC}) on $(x,y)$ plane (a) 
and 
the lines on which their counterparts (\ref{AnalContBC3}) are imposed (b). Figure (c) gives an 
example of such lines which correspond to $AdS_5\times S^5$.
} \label{FigAltBC}
\end{figure}

If we start from the system (\ref{AnlCntD1})--(\ref{AnalContTheEqn}) and look for 
regular geometries, we
 should not allow the AdS 
space to shrink to zero size, and since the dilaton should also stay finite, we 
conclude that on the entire line 
$y=0$ it is $S^2$ that shrinks to zero size. As we know from solving the original
 system, it is impossible 
to have a nontrivial solution unless radius of $S^4$ also goes to zero at come 
points and in the new 
description this should happen somewhere in the upper half of the plane (i.e. at 
$y>0$). Looking at the 
equation (\ref{AnalContTheEqn}) we observe the behavior of terms that do not 
contain $f_1'$:
\bea\label{AnalContBC1}
S^4\quad\mbox{shrinks}:&&y=0,\quad \log\frac{e^{G'}-1}{e^{G'}+1}=0\\
S^2\quad\mbox{shrinks}:&&\arctan\frac{e^{A'}}{F'}=0\nonumber
\eea
To take into account the flux, we decompose it as in (\ref{DefPsi}) and define the 
harmonic function $\Phi'$ as in (\ref{MastHarmonic}):
\bea
(\d_{x'}^2+\d_{y'}^2)\Phi'=0:\quad  
\frac{1}{2}\log\frac{e^{G'}-1}{e^{G'}+1}+\Psi'_1=\d_{y'}\Phi',\quad
\arctan\frac{e^{A'}}{F'}+\Psi'_2=\d_{x'}\Phi'
\eea
Since functions $\Psi'_1$, $\Psi'_2$ are defined only up to one harmonic function,
 we can choose this 
function in a way which makes the boundary conditions (\ref{AnalContBC1}) 
especially convenient
\bea\label{AnalContBC2}
S^4\quad\mbox{shrinks}:&&y'=0,\quad \d_{y'}\Phi'=0,\quad \Psi'_1=0\nonumber\\
S^2\quad\mbox{shrinks}:&&f(x',y')=0,\quad \d_x\Phi'=0,\quad \Psi'_2=0
\eea
Now we have to find the restriction on a curve $f(x',y')=0$. Let us consider some 
point on this curve where 
$y'\ne 0$ (otherwise, both $S^2$ and $S^4$ shrink to to zero and such special 
points require a separate consideration), then we can rewrite equations (\ref{AnlCntD1}) as
\bea
&&\d_{y'} e^{A'}=e^{A'}\left[-\d_{y'}(C'-\frac{\phi}{2})-\frac{e^{2B'}}{y'(e^{2C'}-e^{2B'})}
\right],\nonumber\\
&&\d_{x'} e^{A'}=-e^{A'}\d_{y'}(C'-\frac{\phi}{2})-\frac{ F'}{y'(e^{2C'}-e^{2B'})}
\eea
Since we are considering the point where $e^{A'}=0$, these two equations imply the
 the gradient of 
$e^{A'}$ points along $x'$ direction. In other words, the curves of $e^{A'}=0$ are
 located at the fixed value of $x'$. Then we have to impose the boundary conditions
 along the straight lines depicted in figure \ref{FigAltBC}b:
\bea\label{AnalContBC3}
S^4\quad\mbox{shrinks}:&&y'=0,\quad \d_{y'}\Phi'=0,\quad \Psi'_1=0\nonumber\\
S^2\quad\mbox{shrinks}:&&x'=x_i,\quad 0<y'<y_i,\quad \d_{x'}\Phi'=0,\quad \Psi'_2=0
\eea
Thus the solution is parameterized by a set of pairs $(x_i,y_i)$, then one has to 
solve the Laplace 
equation with boundary conditions (\ref{AnalContBC3}). The resulting harmonic 
function $\Phi'$ leads to 
a unique geometry which is guaranteed to be regular. 


As in section \ref{SectAdS}, we can show that setting three--form to zero, we end 
up with a unique 
solution, which describes $AdS_5\times S^5$. Rather than repeating those arguments
 here, we just state
the result that $AdS_5\times S^5$ corresponds to the boundary conditions along the 
curve depicted in 
figure \ref{FigAltBC}c. Notice that the junction of the vertical and horizontal lines 
is universal (one can 
rescale the coordinates to zoom in on this point), and $AdS_5\times S^5$ example 
demonstrates that no additional singularity develops at such 
junction\footnote{It is interesting to observe that the boundary conditions on 
surfaces similar to ones depicted in figure \ref{FigAltBC}b were encountered in a 
description of 
BPS geometries in M theory \cite{LinMald}. However, unlike the present case which 
has simple 
Neumann boundary conditions on such curves, the boundary conditions discussed in 
\cite{LinMald} were 
more complicated (in that case there was one more coordinate and the "lines" 
$x=x_i$ actually represented the disks).}.

To summarize, we showed that performing a double analytic continuation (\ref{AnalSS}) 
which exchanges $AdS_2$ and $S^2$, one arrives at an alternative description of 
the same system which uses a different 
coordinate frame. However in that frame one can also formulate very simple 
boundary conditions 
(\ref{AnalContBC3}), so we have two equivalent ways of looking at geometries 
with 
$AdS_2\times S^2\times S^4$ factors. We now discuss another analytic continuation 
which leads to a description of new geometries. 

In the geometry (\ref{MastEqnMetric}) we have two spheres, so one can perform one more 
continuation:
\bea
\begin{array}{c}
ds^2_{AdS_2}\rightarrow -ds^2_{S^2}\\ ds^2_{S^4}\rightarrow -ds^2_{AdS_4}
\end{array},
\quad \begin{array}{c}x\rightarrow ix'\\ y\rightarrow iy'\end{array},\quad
G\rightarrow G'-\frac{\pi i}{2},\quad
H\rightarrow H'+\frac{\pi i}{2},\quad 
\begin{array}{c}f_1\rightarrow if_1'\\f_3\rightarrow if_3'
\end{array}
\eea
The resulting geometry is governed by the equations:
\bea
&&d(H'-G'-2\phi)=-\frac{2}{y'(e^{2B'}-e^{2C'})}(-e^{2C'}dy'+ 
F' e^{B'+C'-A'}dx')\\
&&-\frac{1}{2}e^{-2A'-\phi/2}df'_1=\frac{e^{-\phi/2}}{e^{2A'}+e^{2B'}}\left[
e^{A'+\phi/2} F'd\phi-y'*d\phi\right],
\nonumber\\
&&\frac{1}{2}e^{-2B'+\phi/2}df_2=\frac{1}{e^{2A'}+e^{2B'}}\left[-e^{B'} 
Fd\phi-e^{A'+C'}*d\phi\right]\nonumber\\
&&e^{B'} e^{-4C'}*df'_3=e^{A'} d(A'-\frac{\phi}{4})-
\frac{1}{4}F' e^{-\phi/2-2A'}df'_1\nonumber\\
&&*d\arctan (-ie^{G'})+\frac{1}{2}d\log\frac{ie^{A'}-F'}{ie^{A'}+F'}+
\frac{i}{2}e^{-\phi/2-2A'}df'_1=0\\
&&F'\equiv\sqrt{e^{2C'}-e^{2B'}-e^{2A'}}
\eea
For the reference we also give a complete set of SUGRA fields for this case:
\bea
&&ds^2=e^{2A'}(d\Omega'_2)^2+e^{2B'}d\Omega_2^2+e^{2C'}(dH'_4)^2+
\frac{e^{-\phi}}{e^{2C'}-e^{2B'}}
((dx')^2+(dy')^2)\nonumber\\
&&F_5=df'_3\wedge dH'_4+*_{10}(df'_3\wedge dH'_4),\quad
H_3=df'_1\wedge d\Omega'_2,\quad F_3=df_2\wedge d\Omega_2\\
&&e^{2A'}=y'e^{H'-\phi/2},\quad e^{2B'}=y'e^{G'-\phi/2},\quad e^{2C'}=y'e^{-G'-\phi/2},
\nonumber
\eea
This time the harmonic function $\Phi'$ is defined by
\bea
(\d_{x'}^2+\d_{y'}^2)\Phi'=0:\quad  
\frac{1}{2}\log\frac{1-e^{G'}}{e^{G'}+1}+\Psi'_1=\d_{y'}\Phi',\quad
-\arctan\frac{e^{A'}}{F'}+\Psi'_2=\d_{x'}\Phi'
\eea
Notice that for this continuation we again have to impose the boundary conditions 
at $y'=0$ (where one 
of the $S^2$'s goes to zero size), and on certain lines $x'=x_0$, $y'>y_0$ similar to
what we had in 
(\ref{AnalContBC3}):
\bea\label{AnalContBC4}
S^2\quad\mbox{shrinks}:&&y'=0,\quad \d_{y'}\Phi'=0,\quad \Psi'_1=0\nonumber\\
{\tilde S^2}\quad\mbox{shrinks}:&&x'=x_i,\quad y'>y_i>0,\quad \d_{x'}\Phi'=0,\quad 
\Psi'_2=0
\eea
Of course, there is an alternative way of solving the system with $AdS_4\times 
S^2\times S^2$ factors
which is based on introducing a more convenient coordinate 
${\tilde y}=e^{A'+B'-\phi'/2}$, but we will not 
explore this further. 


\section{Back to the brane probes.}
\renewcommand{\theequation}{7.\arabic{equation}}
\setcounter{equation}{0}

\label{SectGoBack}

As we showed in section \ref{SectSumma}, to find geometries with 
$AdS_2\times S^2\times S^4$ factors, one needs to solve a system 
(\ref{MastEqnMetric})--(\ref{MastEqnD2}). Although we were not able to find new 
nontrivial solutions
 of this system, we demonstrated that for the spaces which asymptote to 
$AdS_5\times S^5$, the geometries are uniquely parameterized by one  
harmonic function $\Phi$. Now we want to study some qualitative properties of the 
solutions and show that they are in a perfect agreement with expectations from the
 brane probe analysis which was presented in section \ref{SectDBI}. 

\begin{figure}
\epsfxsize=5.5in \epsffile{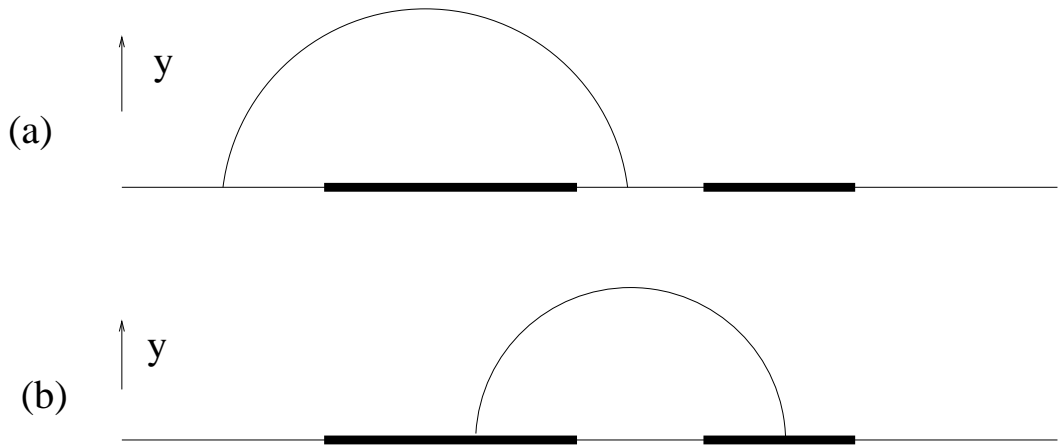}
\caption{The geometries described in this paper have non--trivial topologies which are 
characterized by non--contractible 3-- and 5--cycles. To construct a five--cycle, one looks at a 
contour depicted in figure (a) and fibrates $S^4$ over it. The three--cycles are constructed in a 
similar ways using contours from figure b and  $S^2$.
} \label{FigTopol}
\end{figure}

We begin with discussing the topology of the solutions. Let us consider a generic 
boundary condition depicted in figure \ref{FigDivLine}. In the light region, the $S^4$ 
shrinks to zero size, so it is useful to take a 
contour depicted in  \ref{FigTopol}a and construct a five dimensional manifold as a 
warped product of this contour and $S^4$. Restricting metric (\ref{MastEqnMetric}) to this 
manifold, we find that as $y$ approaches zero, 
the volume of $S^4$ goes to zero as well and near such points metric is 
approximated by
\bea
ds_5^2={\cal F}(dy^2+y^2d\Omega_4^2)
\eea
and it looks like a north pole of $S^5$. We conclude that the five manifold that 
we just described has a topology of $S^5$, moreover if there is a dark region 
between the two endpoints of the contour, (as in figure \ref{FigTopol}a), this $S^5$ 
is not contractible. This is analogous to $S^5$ which emerged in \cite{llm} by 
combining three dimensional sphere and a certain two--dimensional surface. 
Moreover, for the $AdS_5\times S^5$ solution, the five sphere which we 
described here and the one discussed in 
\cite{llm} are the same, and they simply correspond to the $S^5$ factor in the 
geometry. 

Since we have a non--contractible five--manifold and there is a nontrivial 
five--form field strength, it leads to a non--zero flux over such manifold.
We see that the dark strips on the $y=0$ line serve as sources of D3 branes, and 
by the symmetry arguments one can see that these branes have $AdS_2\times S^2$ 
worldvolume. Then we conclude 
that the dark strips describe a gravitational backreaction of D3 branes which were
 discussed in section
\ref{SectDBI}. 


A similar analysis can be performed for contours which end in the dark regions (see figure 
\ref{FigTopol}b). In this case we take a contour and fibrate $S^2$ over it, then we 
arrive at a non--contractible three--manifold which has a topology of $S^3$. Since 
we have a magnetic RR three--form, it can have a non--zero flux over such 
manifold. Then we conclude that the light strips describe gravity solutions for 
the polarized D5 branes which were discussed in section \ref{SectDBI}. Of course, 
the language of D3 branes and D5 branes is 
only appropriate when we have a small dark strip inside a long light region or 
vice versa, otherwise one has a background were all fluxes are turned on and are 
comparable in strength. As far as topology of the solution is concerned, we conclude that it is 
completely determined by the topology of the dark regions on $y=0$ line, and thus it is uniquely 
specified by the set of non--contractible three-- and five--cycles.

\section{Discussion}
\renewcommand{\theequation}{8.\arabic{equation}}
\setcounter{equation}{0}

While we have a very good understanding of supersymmetric branes in flat space, 
the picture is less 
clear for the branes in curved spacetimes. Starting from the original discovery 
of giant gravitons 
\cite{giant}, 
there was a remarkable progress in understanding branes in AdS spaces 
\cite{skenTayl}, but most of the work was devoted to studying the brane probe 
approximation. Such branes are usually curved and to stabilize their shape, they 
are either moving or have some fluxes on the worldvolume, and the stabilization happens 
via interaction between such fluxes and background RR field. It would be nice to understand 
the geometries produced 
by such curved branes and for the giant gravitons of \cite{giant} this problem 
was solved in \cite{llm}. In this paper 
we looked at another class of $1/2$ BPS branes which are supported by fluxes 
rather than angular momentum and we showed that, as in \cite{llm}, the geometries 
are parameterized by one harmonic function with very simple boundary conditions. 
Unfortunately, to translate this harmonic function into the explicit metric, one 
still has to solve certain differential equations and we showed that such solution
 is unique. This is in sharp contrast to a situation in \cite{llm} where one 
starts form a harmonic function and 
recovers the geometry by simple algebraic manipulations. It would be nice if 
better understanding of
equations (\ref{MastEqnF1})-(\ref{MastHarmonic}) could lead to a similar picture for our 
geometries as well.

In this paper we discussed only the branes which preserve half of supersymmetries 
in 
$AdS_5\times S^5$, just as \cite{llm} dealt with 1/2 BPS states but with different bosonic 
 symmetries. It would be interesting to understand the geometries which preserve 
less supersymmetries, especially since they have a very nice description in the 
brane probe approximation. For example, giant gravitons that preserve $1/4$ and 
$1/8$ of the supersymmetries, are described in terms of the holomorphic surfaces 
\cite{mikhail}. In particular, such giant gravitons still preserve $S^3\times R$ 
symmetry which comes from the $AdS$ part of the geometry, but another $SO(4)$ 
(which was crucial for the construction of \cite{llm})
is broken. Unfortunately the problem of finding the gravity solutions for such branes 
reduces to a complicated equation of the Monge--Ampere type, and it is 
not clear what can be learned from it. However, on the field theory side, the 
interesting progress was made in \cite{newBeren}, where it was argued that metrics
 could arise as semiclassical limit of matrix models. Although so far this 
approach has not led to any explicit solutions, this direction appears to be very 
promising.

Recently, the analog of giant gravitons preserving $8$ supersymmetries was discussed in 
the context of branes which are 
dual Wilson lines \cite{dymar}, and such objects are expected to preserve 
$AdS_2\times S^2$ symmetries. It would be interesting to study the gravitational 
description of such objects in a way similar to the one that we discussed here.

\section*{Acknowledgments}

It is a great pleasure to thank Soo-Jong Rey for collaboration during the early stages 
of this work, and for numerous discussions. I also benefited from conversations with 
Jaume Gomis, Nick Halmagyi and Aki Hashimoto. This work was initiated at KITP in 
Santa Barbara and I want to thank the organizers of the program on Higher Dimensional 
Gravity for hospitality and for financial support via NSF Grant No. PHY99-07949. This 
research was supported in part by DOE grant DE-FG02-90ER40560.

\appendix


\section{Solving gravity equations}
\renewcommand{\theequation}{A.\arabic{equation}}
\setcounter{equation}{0}

\label{AppSUGRA}

The main goal of this paper is to find supersymmetric geometries which contain 
$AdS_2\times S^2\times S^4$ factors. The motivation for doing this was given in 
section \ref{SectDBI}, and in this appendix we give some technical steps which led to 
the final solution (\ref{MastEqnMetric})--(\ref{MastEqnD2}). 


\subsection{Formulation of the problem.}

We are looking for supersymmetric solutions of type IIB supergravity, so we begin 
with summarizing the 
fermionic variations using the standard notation of \cite{schwarz}:
\bea\label{GenSUSYVar}
\delta\la&=&i{\not P}\eps^*-\frac{i}{24}\gamma^{mnp}G_{mnp}\eps=0\nonumber\\
\delta \psi_M&=&(\nabla_M-\frac{i}{2}Q_M)\eps+\frac{i}{480}{\not F}_5\gamma_M\eps+
\frac{1}{96}(-\gamma_M{\not G}-2{\not G}\gamma_M)\eps^*
\eea
Supersymmetry parameter $\eps$ is a complex Weyl spinor ($\Gamma_{11}\eps=-\eps$), 
and the expressions for two vectors $Q_m, P_m$ and a scalar $B$ can be found in 
\cite{schwarz} (see also \cite{granaPolch}). 
Below we will write such expressions for a special case.

Equations (\ref{GenSUSYVar}) give SUSY variations for any bosonic background of 
type IIB SUGRA, but we will need a truncated version of these equations. As argued
 in section \ref{SectSumma}, we are interested in  solutions with vanishing axion 
$C^{(0)}$, this implies that $\tau=ie^{-\phi}$, $Q_\mu=0$, and 
\bea
&&P_\mu=\left(1-\left[\frac{1-e^{-\phi}}{1+e^{-\phi}}\right]^2\right)^{-1}\d_\mu 
\frac{1-e^{-\phi}}{1+e^{-\phi}}=
 \frac{(1+e^{-\phi})^2}{4e^{-\phi}}\d_\mu \frac{2}{1+e^{-\phi}}=\frac{1}{2}
\d_\mu\phi\\
&&B=\frac{1-e^{-\phi}}{1+e^{-\phi}},\quad f^{-2}=
\frac{4e^{-\phi}}{(1+e^{-\phi})^2},\nonumber\\
&&G_3=f(H_3+iF_3-BH_3+iBF_3)=e^{-\phi/2}H_3+ie^{\phi/2}F_3
\eea
Substituting these expressions into (\ref{GenSUSYVar}), we arrive at the equations
 which will be analyzed in the remaining part of this appendix:
\bea\label{TheSUSYVar}
&&\delta\la=\frac{i}{2}{\not\d}\phi\eps^*-\frac{i}{24}\gamma^{mnp}G_{mnp}\eps=0\\
&&\delta\psi_M=\nabla_M\eps+\frac{i}{480}{\not F}_5\gamma_M\eps+\frac{1}{96}
(-\gamma_M
{\not G}-2{\not G}\gamma_M)\eps^*=0
\eea
The metric and fluxes are given by equations (\ref{Metric0}), (\ref{3Form0}) and it might 
be useful to reproduce them here:
\bea\label{MetricApp}
ds^2&=&e^{2A}dH_2^2+e^{2B}d\Omega_2^2+e^{2C}d\Omega_4^2+h_{ij} dx^i dx^j\\
F_5&=&df_3\wedge d\Omega_4+*_{10}(df_3\wedge d\Omega_4),
\quad H_3=df_1\wedge dH_2,\quad F_3=df_2\wedge
 d\Omega_2,\quad e^{\phi}
\eea

Equations (\ref{MetricApp}) guarantee that all bosonic fields have the required 
symmetry, but we also need to impose the symmetry on the spinor $\eps$. To do this
 we need to review a construction of spinors on even--dimensional spheres (spinors
 on AdS are trivial modifications of those) and we devote Appendix \ref{AppSpinor}
 to such review. Here we just summarize the results. Let us look at a covariant 
derivative 
$\nabla_m$ along one of the directions of $S^2$ and rewrite it in terms of 
covariant derivative 
${\tilde\nabla}_m$ on a unit two--sphere: 
\bea
\nabla_m\eps={\tilde\nabla}_m\eps-\frac{1}{2}{\gamma^\mu}_m\d_\mu B
\eea
In the appendix \ref{AppSpinor} it is shown that the derivative on a unit sphere 
can be written in terms of hermitean matrix $P_S$ which anticommutes with 
chirality operator on $S^2$ and with gamma matrices along the direction orthogonal
 to this sphere\footnote{In that appendix we always considered reduced gamma 
matrices ${\tilde\gamma}_m$, while here we are writing the ten--dimensional ones. 
This explains an extra factor of $e^{-B}$ in (\ref{EqnApp101}) compared to 
(\ref{SpinMaster}).}:
\bea\label{EqnApp101}
{\tilde\nabla}_m\eps=-\frac{i}{2}e^{-B}\gamma_m P_S\eps
\eea
We can now write the complete derivative of the spinor along $S^2$ direction as 
well as derivatives along $S^4$ and $AdS_2$:
\bea
S^2:&&\nabla_m=-\frac{1}{2}\gamma_m(ie^{-B}P_S-{\not\d}B)\nonumber\\
AdS_2:&&\nabla_m=-\frac{1}{2}\gamma_m(-e^{-A}P_H-{\not\d}A),\\
S^4:&&\nabla_m=-\frac{1}{2}\gamma_m(ie^{-C}P_\Omega-{\not\d}C)\nonumber
\eea
The final ingredient which is needed to write down the equations is the 
expressions for the fluxes. Looking at the formula for the five form flux and 
using projection $\Gamma_{11}\eps=-\eps$, we observe that ${\not F}_5\eps$ can be 
expressed in terms of $f_3$ and we don't have to evaluate the dual piece: 
\bea
&&{\not F}_5\eps=2\times 5!\times e^{-4C}{\not\d}f_3\Gamma_{\Omega}\eps,\quad
\frac{1}{480}{\not F}_5\eps=\frac{e^{-4C}}{2}{\not\d}f_3\Gamma_\Omega\eps
\eea
Here $\Gamma_\Omega$ is a hermitean chirality matrix on $S^4$, and we also 
introduce analogous matrices on $S^2$ (calling it $\Gamma_S$) and $AdS_2$ (it 
will be denoted $\Gamma_H$). Notice that 
the equations (\ref{GenSUSYVar}) are formulated in a basis where all gamma matrices are 
real, this implies that 
$\Gamma_S$ is imaginary, while $\Gamma_H,\Gamma_\Omega$ are real. While we will 
not use an explicit form of gamma matrices, their reality and symmetry properties 
will be important. And rather than summarizing these properties in words, we write
 an explicit basis of gamma matrices which satisfies all the requirements, so the 
reader can consult this equation:
\bea\label{BssGmMtr}
&&P_H=\sigma_2\otimes\sigma_2\otimes 1_4,\quad
P_S=\sigma_2\otimes\sigma_3\otimes\sigma_3\otimes 1_2,\quad
P_\Omega=\sigma_2\otimes\sigma_3\otimes\sigma_2\otimes \sigma_2\nonumber\\
&&\Gamma_H=1_2\otimes\sigma_3\otimes 1_4,\quad
\Gamma_S=1_4\otimes\sigma_2\otimes 1_2,\quad
\Gamma_\Omega=1_8\otimes\sigma_3,\quad
\Gamma_{1,2}=\sigma_{3,1}\otimes 1_8\nonumber\\
&&\Gamma_{11}=-i\Gamma_1\Gamma_2 \Gamma_\Omega\Gamma_S\Gamma_H=
\sigma_2\otimes\sigma_3\otimes\sigma_2\otimes\sigma_3
\eea

A straightforward computation leads to the expression for the three--form flux:
\bea
&&\frac{1}{24}{\not G}=-\frac{1}{4}(e^{\phi/2-2B}{\not\d}f_2 \Gamma_S+
e^{-\phi/2-2A}{\not\d}f_1\Gamma_H)
\eea
The first term in the right hand side of this equation is pure imaginary, while 
the second term is real. Since at some point we will need to do complex 
conjugation, it is useful to define two matrices\footnote{We use the following sign 
conventions for $\Gamma_S$, $\Gamma_H$, $\Gamma_\Omega$. If 
$\Gamma^0$, $\Gamma^9$ are 
gamma matrices corresponding to $AdS_2$ factor, $\Gamma^7$, $\Gamma^8$ are 
matrices corresponding to $S^2$ and $\Gamma^3,\dots,\Gamma^6$ are the ones for 
$S^4$, then we define $\Gamma_H=-\Gamma^0\Gamma^9$,
$\Gamma_S=-i\Gamma^7\Gamma^8$, 
$\Gamma_\Omega=\Gamma^3\Gamma^4\Gamma^5\Gamma^6$. Also thorough this 
paper we use $\Gamma$ to denote matrices with frame indices 
(which square to $\pm 1$), and $\gamma$ stands for 
the matrices with spacetime indices.}:
\bea
G_+=-\frac{1}{4}e^{-\phi/2-2A}{\not\d}f_1\Gamma_H,\quad G_-=-\frac{1}{4}
e^{\phi/2-2B}{\not\d}f_2\Gamma_S,\quad
(G_\pm)^*= \pm G_\pm
\eea
and express the three form flux in terms of them:
\bea
\frac{1}{24}{\not G}=G_++G_-
\eea

Using all this information, we arrive at the final set of equations:
\bea\label{App1eqnPhi}
&&\frac{1}{2}{\not\d}\phi\eps^*-(G_++G_-)\eps=0,\qquad 
\frac{1}{2}{\not\d}\phi\eps-(G_+-G_-)\eps^*=0\\
\label{App1eqnA}
&&(e^{-A}P_H+{\not\d}A)\eps-ie^{-4C}{\not\d}f_3\Gamma_{\Omega}\eps+\frac{1}{2}
(-3G_++G_-)\eps^*=0\\
\label{App1eqnB}
&&(-ie^{-B}P_S+{\not\d}B)\eps-ie^{-4C}{\not\d}f_3\Gamma_{\Omega}\eps+\frac{1}{2}
(G_+-3G_-)\eps^*=0\\
\label{App1eqnC}
&&(-ie^{-C}P_\Omega+{\not\d}C)\eps+ie^{-4C}{\not\d}f_3\Gamma_{\Omega}\eps+
\frac{1}{2}(G_++G_-)\eps^*=0\\
\label{App1eqnDiff}
&&\nabla_\mu\eps+i\frac{e^{-4C}}{2}{\not\d}f_3\gamma_\mu\Gamma_\Omega\eps+
\frac{1}{96}(\gamma_\mu
{\not G}-2\{{\not G},\gamma_\mu\})\eps^*=0
\eea
Notice that the second equation in (\ref{App1eqnPhi}) is just a complex conjugate 
of the first one, but for future reference it is convenient to keep them together.
 It would also be useful to write a hermitean conjugate of the last system:
\bea
&&\frac{1}{2}\eps^T{\not\d}\phi-\eps^\dagger(G_++G_-)=0,\qquad 
\frac{1}{2}\eps^\dagger{\not\d}\phi-
\eps^T(G_+-G_-)=0\\
&&\eps^\dagger(e^{-A}P_H+{\not\d}A)+ie^{-4C}\eps^\dagger{\not\d}f_3\Gamma_{\Omega}
+
\frac{1}{2}\eps^T(-3G_++G_-)=0\\
&&\eps^\dagger(ie^{-B}P_S+{\not\d}B)+ie^{-4C}\eps^\dagger{\not\d}f_3
\Gamma_{\Omega}+
\frac{1}{2}\eps^T(G_+-3G_-)=0\\
&&\eps^\dagger(ie^{-C}P_\Omega+{\not\d}C)-ie^{-4C}\eps^\dagger{\not\d}f_3
\Gamma_{\Omega}+
\frac{1}{2}\eps^T(G_++G_-)=0\\
\label{App1EDCnj}
&&\nabla_\mu\eps^\dagger-i\frac{e^{-4C}}{2}\eps^\dagger\gamma_\mu{\not\d}f_3
\Gamma_\Omega+
\frac{1}{96}\eps^T({\not G}^\dagger\gamma_\mu-2\{{\not G}^\dagger,\gamma_\mu\})=0
\eea
To summarize, we showed that the problem of finding supersymmetric solution with 
metric and fluxes (\ref{MetricApp}) reduces to solving the system 
(\ref{App1eqnPhi})--(\ref{App1eqnDiff}). In the remaining part of this appendix we 
will simplify this 
system and show that its solutions can be parameterized in terms of one harmonic 
function.  


\subsection{Choosing coordinates and evaluating the metric.}


Before we start solving differential equations, it is useful to recall that metric
 (\ref{MetricApp}) is invariant under reparameterizations of $x_1,x_2$ plane, and one can 
use this symmetry to choose a convenient coordinate system.
We begin with adding equations (\ref{App1eqnB}),  (\ref{App1eqnC}) and the second 
equation in (\ref{App1eqnPhi}):
\bea\label{App2proj1}
\left[-ie^{-B}P_S-ie^{-C}P_\Omega+{\not\d}(B+C+\frac{\phi}{2})\right]\eps=0
\eea
This is a projector which generically contains four gamma matrices, but by 
appropriate choice of coordinates and vielbein, we can express it in terms of 
three matrices. Namely we define a coordinate $y$ by a relation 
$y=e^{B+C+\phi/2}$, then in two dimensions we can always choose another coordinate 
$x$ to be orthogonal to $y$. Then metric becomes diagonal and we choose a 
convenient vielbein:
\bea
g_{ij}dx^idx^j=g^2(dy^2+h^2 dx^2),\quad e_y^{\hat y}=g,\quad e_x^{\hat x}=gh
\eea
In this coordinate frame the relation (\ref{App2proj1}) becomes
\bea
\left[-ie^{-B}P_S-ie^{-C}P_\Omega+\frac{1}{yg}\Gamma_y\right]\eps=0
\eea
We already related the product of the warp factors of the spheres with the value 
of coordinate $y$, now it is convenient to parameterize their ratio by function 
$G$. Then the condition for the last equation to be a projector leads to an 
expression for $g$ in terms of $G$:
\bea
e^{2B}=ye^{-\phi/2+G},\quad e^{2C}=ye^{-\phi/2-G},\quad g^2=
\frac{e^{-\phi/2}}{2y\cosh G}
\eea
The projector itself can also be expressed in terms of $G$:
\bea
\left[ie^{-G/2}P_S+ie^{G/2}P_\Omega-\sqrt{e^{G}+e^{-G}}\Gamma_y\right]\eps=0
\eea
and this expression can be further simplified by introducing a rescaled spinor 
$\eps_1$:
\bea\label{DefineEps1}
\eps=e^{-\delta P_\Omega P_S}\eps_1,\qquad (iP_\Omega-\Gamma_y)\eps_1=0,\quad 
\cos 2\delta=\frac{e^{G/2}}{\sqrt{e^{G}+e^{-G}}}
\eea
Once this projection is imposed, the equations (\ref{App1eqnB}),  (\ref{App1eqnC})
 and (\ref{App1eqnPhi}) become linearly dependent and we can disregard equation 
(\ref{App1eqnB}). 

There is one more more combination of 
(\ref{App1eqnPhi})---(\ref{App1eqnC}) which does not contain fluxes: adding 
(\ref{App1eqnA}) and (\ref{App1eqnC}) and subtracting (\ref{App1eqnPhi}), we find 
a projector\footnote{In this paper we encounter numerous trigonometric and 
hyperbolic functions of various arguments. To avoid writing formulas which are 
unnecessarily long, we adopt a shorthand notation: $$s_x\equiv \sin x, \quad 
c_x\equiv \cos x,\quad sh_x\equiv \sinh_x,\quad ch_x\equiv \cosh x$$.}
\bea
&&(e^{-A}P_H-ie^{-C}P_\Omega+{\not\d}(A+C-\frac{\phi}{2}))\eps=0:\nonumber\\
&&\qquad\qquad\qquad(e^{-A}P_H-ie^{-C}(c_\delta P_\Omega-s_\delta P_S)+
{\not\d}(A+C-\frac{\phi}{2}))\eps_1=0
\eea
We wrote this equation in terms of $\eps_1$ because this spinor satisfies a very 
simple projection relation (\ref{DefineEps1}). In particular, acting on the last 
equation by $(\Gamma_y\pm iP_\Omega)$, we find two relations:
\bea
(e^{-A}P_H+ie^{-C}s_\delta P_S+\frac{1}{gh}\Gamma^x\d_x(A+C-\frac{\phi}{2}))\eps_1=0,\quad
(-e^{-C}c_\delta+\frac{1}{g}{\d}_y(A+C-\frac{\phi}{2}))\eps_1=0\nonumber
\eea
The first equation suggests that it is convenient to rescale a spinor one more 
time and to define a useful function $F$:
\bea\label{defEps0E1}
\eps_1=e^{i\sigma P_S P_H}\eps_0,\quad \tanh 2\sigma=e^{-A}\sqrt{e^{2B}+e^{2C}},
\quad
F=\sqrt{e^{2A}-e^{2B}-e^{2C}}
\eea
This leads to a simple expressions for the derivatives of $A+C-\phi$:
\bea\label{defEps0E2}
{\d}_y(A+C-\frac{\phi}{2})=\frac{e^G}{y(e^G+e^{-G})},\quad 
\d_x(A+C-\frac{\phi}{2})=-\frac{\alpha hFe^{-A}}{y(e^{G}+e^{-G})}
\eea
and to a simple projection relation
\bea\label{defEps0E3}
\left[iP_S-\alpha\Gamma_x\right]\eps_0=0
\eea
Here $\alpha$ is a parameter which is equal to plus or minus one, and we will fix 
its value later.  As before, we conclude that the system (\ref{defEps0E2}), 
(\ref{defEps0E3}) can be viewed as a replacement for the equation 
(\ref{App1eqnC}).

At this point the metric is still invariant under reparameterizations of $x$ and 
it would be nice to find a convenient gauge. Under such reparameterizations, it 
is only function $h$ which changes, so to fix the gauge we will need to know the 
$y$--dependence of $h$. The simplest way to address this question is to look at 
certain spinor bilinears and find the differential equations for them. Notice, 
that in principle to find a supersymmetric background it is sufficient to analyze 
all spinor bilinears, and this technique was very fruitful in the recent years 
\cite{gaunt}. In particular, in 
\cite{llm} it was used to find 1/2 BPS geometries with $SO(4)\times SO(4)$ 
symmetry which is analogous to the problem which we are considering. However in 
this paper we mostly work with spinors directly, and one of the reason for this 
is that one can construct many bilinears by placing matrices 
$P_H$, $P_S$, $P_\Omega$, $\Gamma_H$, $\Gamma_S$, $P_\Omega$, $\gamma_{1,2}$ 
between spinors and many of these bilinears turn out to be zero. Also some of the 
bilinears in the set are equal to others, so it appears that the exhaustive 
analysis of bilinears in the present case would be longer than a direct search for
 a solution of spinor equations. However we will now use some of the bilinears to 
determine the function $h$.

Since we want to find a differential equation for $e^{\hat x}_x$, it is natural to
 start from a vector bilinear which has only a component along $x$ direction. The 
projection (\ref{DefineEps1}) shows that one such bilinear is\footnote{We are using conventions 
$\Gamma_x\Gamma_y=i{\hat\sigma}_2$.}
\bea
&&\eps^\dagger \Gamma_\Omega\gamma_\mu P_\Omega\eps=\eps_1^\dagger \Gamma_\Omega
\gamma_\mu P_\Omega\eps_1:\quad
\eps^\dagger \Gamma_\Omega\gamma_y P_\Omega\eps=0,\quad
\eps^\dagger \Gamma_\Omega\gamma_x P_\Omega\eps=e_x^{\hat x}\eps_1^\dagger
\Gamma_\Omega
{\hat\sigma}_2\eps_1\nonumber
\eea
One--form constructed from this vector can be expressed in terms of a scalar 
bilinear: 
\bea
\eps^\dagger \Gamma_\Omega\gamma_\mu P_\Omega\eps~ dx^\mu=
e_x^{\hat x}e^y_{\hat y}e^{-\phi/2-B} \eps^\dagger\Gamma_\Omega{\hat\sigma}_2
\eps~dx=
h e^{-\phi/2-B} \eps^\dagger\Gamma_\Omega{\hat\sigma}_2\eps~dx
\eea
Knowing an exterior derivative of this vector as well as coordinate dependence of 
the scalar bilinear, we can extract a $y$--dependence of $h$. We begin with 
computation of the exterior derivative using (\ref{App1eqnDiff}):
\bea
&&\nabla_\mu(\eps^\dagger \Gamma_\Omega\gamma_\nu P_\Omega\eps)+\frac{i}{2}e^{-4C}
\left(\eps^\dagger \Gamma_\Omega\gamma_\nu P_\Omega{\not\d}f_3\gamma_\mu
\Gamma_\Omega\eps-
hc\right)\nonumber\\
&&\qquad-\frac{1}{4}\left[\eps^\dagger \Gamma_\Omega\gamma_\nu P_\Omega
(\gamma_\mu(G_++G_-)+2(G_++G_-)\gamma_\mu)\eps^*+hc\right]=0\nonumber
\eea
Using (\ref{App1eqnPhi}) to exclude $G_-\eps^*$ from this expression, and taking 
antisymmetric part in $\mu,\nu$ indices, we find
\bea
\nabla_{[\mu}(\eps^\dagger \Gamma_\Omega\gamma_{\nu]} P_\Omega\eps)
+\frac{1}{4}\left[\eps^\dagger \Gamma_\Omega P_\Omega\left(\gamma_{\nu\mu} 
(-\frac{1}{2}{\not\d}\phi\eps+2G_+\eps^*)+
2\gamma_{[\nu}(G_++G_-)\gamma_{\mu]}\eps^*\right)+hc\right]=0\nonumber
\eea
Noticing that
\bea
\Gamma_x \Gamma_\la\Gamma_y-\Gamma_y \Gamma_\la\Gamma_x=2\delta_\la^x(\Gamma_y-
\Gamma_y)-\Gamma_\la \Gamma_{xy}-\Gamma_{xy}\Gamma_\la=0\nonumber
\eea
we simplify the equation above:
\bea
&&\nabla_{[\mu}(\eps^\dagger \Gamma_\Omega\gamma_{\nu]} P_\Omega\eps)
-\frac{i}{4}\eps_{\mu\nu}\left[\eps^\dagger \Gamma_\Omega P_\Omega{\hat\sigma}_2 
(-\frac{1}{2}{\not\d}\phi\eps+2G_+\eps^*)+hc\right]=0
\eea
Now one can use the explicit form of $G_+$ along with relation $\Gamma_H=
\Gamma_H^T$ to evaluate the transpose of the term involving field strength: 
\bea
(\eps^\dagger \Gamma_\Omega P_\Omega{\hat\sigma}_2 G_+\eps^*)^T=
-\eps^\dagger \Gamma_\Omega P_\Omega{\hat\sigma}_2 G_+\eps^*=0
\eea
Then finally get an equation
\bea
&&\nabla_{[\mu}(\eps^\dagger \Gamma_\Omega\gamma_{\nu]} P_\Omega\eps)
+\frac{i}{4}\eps_{\mu\nu}\eps^\dagger \Gamma_\Omega P_\Omega{\hat\sigma}_2{\not\d}
\phi\eps=0\nonumber
\eea
Writing this in terms of forms, and taking a coefficient in front of 
$dy\wedge dx$, we find:
\bea\label{EqnForH}
\d_y(h e^{-\phi/2-B} \eps^\dagger\Gamma_\Omega{\hat\sigma}_2\eps)+
\frac{1}{2}h e^{-\phi/2-B} \eps^\dagger\Gamma_\Omega{\hat\sigma}_2\eps~\d_y\phi=0.
\eea
To extract a $y$--dependence of $h$ we need to know a functional form of the 
bilinear appearing in this relation. Starting from differential equations 
(\ref{App1eqnDiff}), (\ref{App1EDCnj}) one can write an expression for the derivative of 
this bilinear, 
then using (\ref{App1eqnPhi}) to remove $G_+$ from the result, one arrives at
\bea\label{DiffBilin1}
\nabla_\mu(\eps^\dagger{\hat\sigma}_2\Gamma_\Omega\eps)+i\frac{e^{-4C}}{2}
\eps^\dagger 
{\hat\sigma}_2[{\not\d}f_3,\gamma_\mu]\eps
-\left(\frac{1}{2}\eps^\dagger {\hat\sigma}_2\Gamma_\Omega\gamma_\mu G_-\eps^*-
\frac{1}{8}\d_\mu\phi \eps^\dagger{\hat\sigma}_2\Gamma_\Omega\eps+hc\right)=0\nonumber\\
\eea
The term involving three--form can be evaluated by looking at combination of 
(\ref{App1eqnPhi}) and 
(\ref{App1eqnB}):
\bea
&&\left[-ie^{-B}P_S+{\not\d}(B+\frac{\phi}{4})-ie^{-4C}{\not\d}f_3\Gamma_{\Omega}
\right]\eps-G_-\eps^*=0
\nonumber
\eea
and at the conjugate relation. This leads to equation
\bea
&&\frac{1}{2}\left(\eps^\dagger {\hat\sigma}_2\Gamma_\Omega\gamma_\mu G_-\eps^*+
hc\right)=
\eps^\dagger {\hat\sigma}_2\Gamma_\Omega
\left[{\d}_\mu(B+\frac{\phi}{4})-\frac{i}{2}e^{-4C}[\gamma_\mu,{\not\d}f_3]
\Gamma_{\Omega}\right]\eps
\eea
Substituting this into the equation for the scalar bilinear, we find a very 
simple relation which can be solved in terms of one integration constant $c_1$:
\bea\label{ScalarBilin}
\nabla_\mu(\eps^\dagger{\hat\sigma}_2\Gamma_\Omega\eps)-
\eps^\dagger{\hat\sigma}_2\Gamma_\Omega\eps~\d_\mu B=0:\qquad
\eps^\dagger{\hat\sigma}_2\Gamma_\Omega\eps=c_1e^{B}
\eea
We will show below that $c_1$ is not equal to zero (see equation (\ref{App3Eqn51})) and 
this fact will not rely on a particular value of $h$. It is only for 
presentational purposes that we postpone the derivation of (\ref{App3Eqn51}) until the 
next subsection. 
Substituting (\ref{ScalarBilin}) into (\ref{EqnForH}) and dividing result by 
non--vanishing $c_1$, we conclude that function $h$ does not depend on $y$, so we 
can choose a gauge where $h=1$.

To summarize, we fixed the diffeomorphism--invariance in the metric, and we shown 
that it can be written in terms of two independent warp factors and the dilaton:
\bea
&&ds^2=e^{2A}dH_2^2+e^{2B}d\Omega_2^2+e^{2C}d\Omega_4^2+\frac{e^{-\phi}}{e^{2B}+
e^{2C}}
(dx^2+dy^2)\\
&&e^{2A}=ye^{H-\phi/2},\quad e^{2B}=ye^{G-\phi/2},\quad e^{2C}=ye^{-G-\phi/2}\nonumber
\eea
Notice that this result was already obtained in \cite{yama}, but to be able to go 
further, we had to re--derive it in the standard notation. We also showed that 
the Killing spinor should satisfy four algebraic relations: (\ref{App1eqnPhi}), 
(\ref{App1eqnA}) and 
\bea\label{DynamProject}
\left[iP_S-\alpha\Gamma_x\right]\eps_0=0,\quad [iP_\Omega-\Gamma_y]\eps_0=0,\quad
\eps=e^{-\delta P_\Omega P_S}e^{i\sigma P_S P_H}\eps_0
\eea
Along with differential equations (\ref{App1eqnDiff}) and (\ref{defEps0E2}), these
 relations give a complete system of equations which we solve in the next 
subsection. We conclude by rewriting the differential equations (\ref{defEps0E2}) 
in terms of $G$ and $H$:
\bea\label{GHeqnApp}
\frac{1}{2}\d_y(H-G-2\phi)=-\frac{e^{-G}}{y(e^G+e^{-G})},\quad
\frac{1}{2}\d_x(H-G-2\phi)=-\frac{\alpha Fe^{-A}}{y(e^G+e^{-G})}
\eea


\subsection{Evaluating the fluxes.}

We begin with looking at the dilatino variation (\ref{App1eqnDiff}) and rewriting 
it in terms of $\eps_0$:
\bea\label{App3Grvtno}
\frac{1}{2}{\not\d}\phi\eps_0^*-
(e^{-2i\sigma P_SP_H}G_++c_{2\delta}e^{-2i\sigma P_SP_H}G_-+s_{2\delta}P_\Omega 
P_S G_-)\eps_0=0
\eea
We want to take various projections of this equation, and it seems convenient to 
write the matrices $G_+$ and $G_-$ in terms of scalars:
\bea
G_+\equiv G_{+,x}\gamma^x\Gamma_H+G_{+,y}\gamma^y\Gamma_H,\quad 
G_-\equiv G_{-,x}\gamma^x\Gamma_S+G_{-,y}\gamma^y\Gamma_S
\eea
Let us act on (\ref{App3Grvtno}) by $\gamma_x(1+i\Gamma_yP_\Omega)$ then using the
 relation 
$P_\Omega P_S\Gamma_y\Gamma_x\eps_1=\alpha\eps_1$, we arrive at the equation
\bea
\frac{1}{2}{\d}_x\phi\eps_0^*-
(e^{-2i\sigma P_SP_H}G_{+,x}\Gamma_H+
c_{2\delta}e^{-2i\sigma P_SP_H}G_{-,x}\Gamma_S+\alpha s_{2\delta}G_{-,y}\Gamma_S)
\eps_0=0
\eea
Projecting this relation by $(1\pm i\alpha \Gamma_x P_S)$, we find :
\bea\label{App2Eqn101}
\frac{1}{2}{\d}_x\phi\eps_0^*-
(ch_{2\sigma}G_{+,x}\Gamma_H-
ic_{2\delta}sh_{2\sigma}P_SP_HG_{-,x}\Gamma_S)\eps_0=0\nonumber\\
-i\Gamma_S P_S \Gamma_HP_H\eps_0~ sh_{2\sigma}G_{+,x}=
(c_{2\delta}ch_{2\sigma}G_{-,x}+\alpha s_{2\delta}G_{-,y})\eps_0
\eea
Assuming the the three--form flux doesn't vanish, we conclude that there is an 
additional projection relation:
\bea\label{App3Eqn102}
i\Gamma_S P_S \Gamma_HP_H\eps_0=\beta \eps_0
\eea 
where $\beta=\pm 1$. Then equations (\ref{App2Eqn101}) can be rewritten as 
\bea\label{App3Eqn104in}
&&c_{2\delta}ch_{2\sigma}G_{-,x}+\alpha s_{2\delta}G_{-,y}+\beta sh_{2\sigma}
G_{+,x}=0\\
&&\frac{1}{2}{\d}_x\phi\Gamma_H\eps_0^*-
(ch_{2\sigma}G_{+,x}+\beta
c_{2\delta}sh_{2\sigma}G_{-,x})\eps_0=0
\eea
Since all coefficients in the last equation are real, for solutions with 
nontrivial dilaton there is one more restriction on 
$\eps_0$: 
\bea\label{App3Eqn103}
\Gamma_H\eps^*_0=a\eps_0,\quad a=\pm 1
\eea
Similarly, acting on (\ref{App3Grvtno}) by $\gamma_y(1-i\Gamma_yP_\Omega)$, 
we find
\bea
&&c_{2\delta}ch_{2\sigma}G_{-,y}-\alpha s_{2\delta}G_{-,x}+\beta sh_{2\sigma}
G_{+,y}=0\\
\label{App3Eqn104out}
&&\frac{1}{2}{\d}_y\phi\Gamma_H\eps_0^*-
(ch_{2\sigma}G_{+,y}+
\beta c_{2\delta}sh_{2\sigma}G_{-,y})\eps_0=0
\eea
Notice that if dilaton is equal to constant, then we get a homogeneous system of 
equations for four components of flux, and the determinant of the appropriate 
matrix is equal to
 $ch^2_{2\sigma}-\frac{1}{2}sh^2_{2\sigma}s_{4\delta}>0$, so 
it we want a solution with nontrivial three--form flux, the dilaton should not 
vanish and projection (\ref{App3Eqn103}) should be enforced. For a vanishing 
three--form flux, the spinor can be chosen to be real, but we can choose a 
modified 
"reality condition" (\ref{App3Eqn103}) as well. Substituting the expressions for 
$G_{\pm,x}$ and $G_{\pm,y}$ in 
(\ref{App3Eqn104in})--(\ref{App3Eqn104out}) and rewriting the result in terms of 
differential forms, we arrive at two equations:
\bea
&&-\beta sh_{2\sigma}e^{-\phi/2-2A}d f_1=
e^{\phi/2-2B}[c_{2\delta}ch_{2\sigma}d f_2+\alpha s_{2\delta}*d f_2]\\
&&ad\phi=-\frac{1}{2}\left[ch_{2\sigma}e^{-\phi/2-2A}d f_1+\beta e^{\phi/2-2B}
c_{2\delta}sh_{2\sigma}d f_2\right]
\eea
Here and below the star represents a Hodge duality in two dimensions with a sign 
convention: $*dy=dx$. The last two relations can be viewed as equations for $df_1$
 and $df_2$ and straightforward algebraic manipulations lead to the solution of 
this system:
\bea\label{DefF12Eqn}
df_1&=&-\frac{2ae^{2A+\phi/2}}{e^{2A}-e^{2B}}\left[e^{A}Fd \phi-\alpha e^{B+C} 
*d\phi\right]\nonumber\\
d f_2&=&2a\beta\frac{e^{-\phi/2+2B}}{e^{2A}-e^{2B}}\left[
e^{B}Fd\phi-\alpha e^{A+C}*d \phi\right]
\eea

Let us pause for a moment and collect all projection relations which have been 
imposed on $\eps_0$ so far. We have (\ref{DynamProject}),  (\ref{App3Eqn102}), 
(\ref{App3Eqn103}) and the standard projector with $\Gamma_{11}$, and these five 
projectors commute with each other. One can also check that these projectors are 
independent (for example, using the explicit basis (\ref{BssGmMtr})), so they 
reduce a dimension of a spinor by a factor of $2^5=32$. Notice that in the basis 
(\ref{BssGmMtr}) we have a 16--component complex spinor, so the 
projections imply that it can be parameterized in terms of one real function\footnote{Of course, the 
spinor in type IIB supergravity has  32 complex components (before the 
$\Gamma_{11}$ projection is imposed), but we suppressed the directions along the 
spheres and AdS. So starting with complex spinor which has one real  component in 
our notation, we produce 
an object which has $2\times 2\times 4=16$ real components. This is expected since
 we are looking at states which preserve $1/2$ of supersymmetries.}.
 This
 explains why we chose to work with spinor directly rather than to write down all 
bilinears following \cite{llm}: to determine the spinor completely we only need 
one real bilinear out of a large set of expressions. In fact we already 
encountered a useful bilinear in (\ref{ScalarBilin}), now we will take a closer look at it.

First we want to show that $c_1$ is a non--vanishing constant. To this end we will
 use various projectors to express the bilinear (\ref{ScalarBilin}) in terms of $\eps_0$: 
\bea\label{App3Eqn51}
\eps^\dagger{\hat\sigma}_2\Gamma_\Omega\eps=c_{2\delta}
\eps^\dagger_0{\hat\sigma}_2\Gamma_\Omega(i~sh_{2\sigma}P_SP_H)\eps_0=
\beta c_{2\delta}sh_{2\sigma}\eps^\dagger_0{\hat\sigma}_2\Gamma_\Omega 
\Gamma_S\Gamma_H\eps_0=
-\beta c_{2\delta}sh_{2\sigma}\eps^\dagger_0\eps_0
\eea
Here we used the definition of $\Gamma_{11}$ as well as projection which it 
imposes: 
\bea
\Gamma_{11}=-i\Gamma_x\Gamma_y \Gamma_\Omega\Gamma_S\Gamma_H=
{\hat\sigma}_2\Gamma_\Omega\Gamma_S\Gamma_H,\qquad \Gamma_{11}\eps=-\eps
\eea
Equation (\ref{App3Eqn51}) implies that unless the Killing spinor $\eps_0$ is 
identically equal to zero, the bilinear 
$\eps^\dagger{\hat\sigma}_2\Gamma_\Omega\eps$ does not vanish, which proves that 
coefficient $c_1$ in (\ref{ScalarBilin}) is not equal to zero\footnote{Although in this 
subsection we already took $h=1$, one can show that all  projection relations 
remain the same for an arbitrary $h$, so to arrive at relation (\ref{App3Eqn51}) 
one does not rely on equation (\ref{ScalarBilin}) (otherwise the logic would be circular).
 We chose to write equations in this order only to avoid unnecessary 
complications.}, then we can rescale a spinor $\eps_0$ to set $c_1=-\beta$. Now 
equation (\ref{DiffBilin1}) can be viewed as a differential equation for $B$, and 
since one bilinear determined the Killing spinor completely, the relation 
(\ref{DiffBilin1}) along with projectors that we discussed is equivalent to 
(\ref{App1eqnDiff}). Substituting the value of $G_-$ into (\ref{DiffBilin1}), we find a 
simple differential equation:
\bea\label{DiffBilin2}
\d_\mu(-\beta e^B)+e^{-4C}\eps^\dagger\eps~ \eps_{\mu\nu}\d_\nu f_3-
\frac{\beta}{4} e^B\d_\mu\phi
+\frac{1}{8}e^{-2B+\phi/2}\d_\mu f_2
\left[\eps^\dagger{\hat\sigma}_2\Gamma_\Omega\Gamma_S\eps^*+cc\right]=0
\eea
Using various projectors, we evaluate the bilinears that appear in this 
expression:
\bea
\eps^\dagger\eps&=&ch_{2\sigma}\eps_0^\dagger\eps_0=\frac{e^A}{F}
\frac{e^B}{c_{2\delta}sh_{2\sigma}}=
\frac{e^A}{F} F=e^A
\\
\eps^T{\hat\sigma}_2\Gamma_\Omega\Gamma_S\eps&=&-\eps_0^T\Gamma_H\eps_0=
-a~ \eps_0^\dagger\eps_0=-aF
\eea
Substituting this into (\ref{DiffBilin2}) and rewriting the result in terms of 
forms, we find the expression for the five--form flux:
\bea\label{DiffBilinB}
\beta e^{-4C+A}*d f_3=e^B d(B+\frac{\phi}{4})+\frac{a\beta}{4}Fe^{-2B+\phi/2}
df_2
\eea
This equation replaces (\ref{App1eqnDiff}). For future reference we write an alternative 
form of the last equation, which can be obtained by combining it with (\ref{defEps0E2}) 
and (\ref{DefF12Eqn}):
\bea\label{DiffBilinA}
\beta e^{-4C+B}*d f_3=e^A d(A-\frac{\phi}{4})+\frac{a}{4}Fe^{-2A-\phi/2}d f_1
\eea
At this point the complete system of bosonic equations is given by (\ref{defEps0E2}), 
(\ref{DefF12Eqn}), (\ref{DiffBilinA}), and in addition we should keep one of the three 
equations 
(\ref{App1eqnA})--(\ref{App1eqnC}). Let us look at (\ref{App1eqnB}) and rewrite it in 
terms of 
$\eps_0$:
\bea\label{AppEqn61}
\left[-ie^{-B}P_S+e^{i\sigma P_SP_H}e^{-2\delta P_\Omega P_S}e^{i\sigma P_SP_H}
{\not\d}(B+\frac{\phi}{4})-
ie^{-4C}e^{2i\sigma P_SP_H}{\not\d}f_3\Gamma_{\Omega}\right]\eps_0-G_-\eps_0^*=0
\nonumber\\
\eea
Here we also used the dilatino equation to eliminate $G_+$. To proceed it is 
useful to combine the projection relations for $\eps_0$ to construct one more 
projector:
\bea
S=\alpha \Gamma_x\Gamma_y P_S P_\Omega:\qquad S\eps_1=\eps_1,\quad S\eps_1^*=
\eps_1^*
\eea
We can now decompose (\ref{AppEqn61}) into two equations by applying $1\pm S$ to 
it. It turns our that after acting by $1+S$ on (\ref{AppEqn61}), we get an 
equation which is equivalent to (\ref{DiffBilinB}), however acting by $1-S$ we 
find a new relation:
\bea
&&\left[-ie^{-B}P_S+(c_{2\delta}ch_{2\sigma}-P_\Omega P_S s_{2\delta} ){\not\d}
(B+\frac{\phi}{4})+
e^{-4C}sh_{2\sigma} P_SP_H {\not\d}f_3\Gamma_{\Omega}\right]\eps_0=0\nonumber\\
&&\left[-ie^{-B}\alpha {\hat\sigma}_2 \Gamma_y+(c_{2\delta}ch_{2\sigma}+
i\alpha{\hat\sigma}_2 s_{2\delta} ){\not\d}(B+\frac{\phi}{4})+
i\beta{\hat\sigma}_2
e^{-4C}sh_{2\sigma} {\not\d}f_3\right]\eps_0=0\nonumber
\eea
Noticing that the first term can be expressed in terms of derivative as 
\bea
-ie^{-B}\alpha {\hat\sigma}_2 \Gamma_y=-i\alpha {\hat\sigma}_2 
\frac{e^{-B-\phi/2}}{\sqrt{e^{2B}+e^{2C}}} {\not\d}y=
-i\alpha {\hat\sigma}_2 s_{2\delta}{\not\d}\log y,
\eea
we find the last projection relation:
\bea\label{BprojWrSgn}
\left[ic_{2\delta}ch_{2\sigma}{\hat\sigma}_2{\not\d}(B+\frac{\phi}{4})
-\alpha s_{2\delta} {\not\d}(B+\frac{\phi}{4}-\log y)+\beta
e^{-4C}sh_{2\sigma} {\not\d}f_3\right]\eps_0=0
\eea
As we mentioned before, at this point $\eps_0$ is essentially a one--component 
spinor, so we cannot impose any more restrictions on it. This implies that in 
(\ref{BprojWrSgn}) the coefficients in front of $\Gamma_x$ and 
$\Gamma_y$  have to vanish separately. An alternative way of seeing this to act 
on (\ref{BprojWrSgn}) by $(1-i\alpha \Gamma_x P_S)$ and to use the projector 
(\ref{defEps0E3}). Thus we end up with 
equation
\bea
\beta e^{-4C}*df_3=\frac{e^{A+B}}{e^{2B}+e^{2C}}d(B+\frac{\phi}{4})
+\alpha \frac{Fe^C}{ e^{2B}+e^{2C}}*d(B+\frac{\phi}{4}-\log y)
\eea
We can exclude five--form flux $f_3$ from the last equation by combining it with 
(\ref{DiffBilinB}), then using 
(\ref{GHeqnApp}) we find the relation
\bea\label{AppHarmEqn}
-\frac{e^{B+C}dG}{e^{2B}+e^{2C}}+\frac{\alpha e^A}{2F}*d\log\frac{e^G+e^{-G}}{e^H}
-\frac{a\alpha}{2} e^{-\phi/2-2A} *df_1=0
\eea
To summarize, we have shown that the system of five differential equations 
(\ref{App1eqnPhi})--(\ref{App1eqnDiff}) can be rewritten 
as four projectors (\ref{DynamProject}), (\ref{App3Eqn102}), (\ref{App3Eqn103}), 
and five differential relations (\ref{GHeqnApp}), (\ref{DefF12Eqn}), (\ref{DiffBilinA}), 
(\ref{AppHarmEqn}) 
and 
these two descriptions are equivalent. 
For future reference, in the next subsection we collect all equations in one 
place.


\subsection{Summarizing supergravity solution.} 

In this long appendix we analyzed the SUSY variations of type IIB supergravity on 
a manifold with 
$AdS_2\times S^2\times S^4$ factors. Let us now collect the results. We showed 
that one can always choose a coordinate system so the the metric and fluxes have 
a form
\bea
&&ds^2=e^{2A}dH_2^2+e^{2B}d\Omega_2^2+e^{2C}d\Omega_4^2+
\frac{e^{-\phi}}{e^{2B}+e^{2C}}
(dx^2+dy^2)\\
&&F_5=df_3\wedge d\Omega_4+*_{10}(df_3\wedge d\Omega_4),\quad
H_3=df_1\wedge dH_2,\quad F_3=df_2\wedge d\Omega_2\\
&&e^{2A}=ye^{H-\phi/2},\quad e^{2B}=ye^{G-\phi/2},\quad e^{2C}=ye^{-G-\phi/2},\quad 
F=\sqrt{e^{2A}-e^{2B}-e^{2C}}\nonumber
\eea
The geometry is supersymmetric if and only is 
these fields satisfy the following differential relations 
\bea
&&df_1=-\frac{2ae^{2A+\phi/2}}{e^{2A}-e^{2B}}\left[e^A Fd\phi-\alpha 
e^{B+C}*d\phi\right],\\
&&df_2=\frac{2a\beta e^{2B-\phi/2}}{e^{2A}-e^{2B}}\left[e^B Fd\phi-\alpha 
e^{A+C}*d\phi\right]\\
&&\beta e^B e^{-4C}*df_3=e^A d(A-\frac{\phi}{4})+\frac{a}{4}F e^{-\phi/2-2A}df_1\\
&&d(H-G-2\phi)=-\frac{2}{y(e^{2B}+e^{2C})}(e^{2C}dy+\alpha F e^{B+C-A}dx)\\
&&\alpha*d\arctan e^G+\frac{1}{2}d\log\frac{e^A-F}{e^A+F}-
\frac{a}{2}e^{-\phi/2-2A}df_1=0
\eea
In this coordinate system we expressed the Killing spinor $\eps$ in terms of a 
reduced spinor $\eps_0$ which effectively has one real component due to 
projections imposed on it\footnote{Nevertheless the solution is 1/2 BPS: see 
footnote 16.}:
\bea
&&\eps=e^{-\delta P_\Omega P_S}e^{i\sigma P_S P_H}\eps_0,\quad 
\tan 2\delta=e^{-G},\quad
\tanh 2\sigma=e^{-A}\sqrt{e^{2B}+e^{2C}}\\
&&\eps_0=-\Gamma_{11}\eps_0=i\Gamma_y P_\Omega\eps_0=i\alpha \Gamma_x P_S\eps_0=
a\Gamma_H\eps_0^*=i\beta \Gamma_S P_S\Gamma_H P_H\eps_0
\eea
We also determined the spinor $\eps_0$ (up to overall normalization) by computing 
its bilinear:
\bea
\eps^\dagger_0\eps_0=\sqrt{e^{2A}-e^{2B}-e^{2C}}
\eea
Each of the constants $a,\alpha,\beta$ can be equal to either plus or minus one, and so far we 
have not fixed their signs. To avoid unnecessary complications, we will fix these projections in the 
main body of the paper by taking $a=\alpha=\beta=1$. This choice does not make the situation 
less general, moreover $a,\alpha,\beta$ can be recovered easily by noticing that the differential relations written in this subsection remain invariant it we flip signs of all elements in any of the following sets:
\bea
(a,f_1,f_2),\quad (\alpha,x),\quad (\beta,f_2,f_3)
\eea
Here $x$ is one of the coordinates which so far was defined as being orthogonal to $y$ and thus 
its sign was not fixed up to this point.

We conclude this summary by writing two useful relations. By combining 
(\ref{DiffBilinA}) and 
(\ref{DiffBilinB}) we arrive at the following equation:
\bea
\frac{1}{2}d[e^{2A}-e^{2B}]-\frac{1}{4}(e^{2A}+e^{2B})d\phi+\frac{a}{2}Fe^A 
e^{-\phi/2-2A}df_1+\frac{1}{2}F^2 d\phi=0
\eea
and (\ref{DiffBilinB}), (\ref{GHeqnApp}), (\ref{AppHarmEqn}) can be combined into
\bea
\alpha e^{B+C} *d(C-\frac{\phi}{4})-\frac{a}{4}e^{2B} e^{-\phi/2-2A}df_1-
F e^A d(A-\frac{\phi}{4})-\frac{a}{4}F^2 e^{-\phi/2-2A}df_1=0
\eea


\section{Constructing spinors on the sphere.}
\renewcommand{\theequation}{B.\arabic{equation}}
\setcounter{equation}{0}

\label{AppSpinor}

While deriving the supersymmetry variations, we encountered spinors on unit 
spheres in even dimensions and in this appendix we summarize a construction of 
such spinors. First we recall that on even--dimensional sphere there are two types
 of Killing spinors, each class satisfies one of the equations \cite{pope}:
\bea
&&\nabla_m\eps^{(1)}_{\pm}=\pm\frac{i}{2}\gamma_m\eps^{(1)}_{\pm},\\
\label{EvenSphereChiral}
&&\nabla_m\eps^{(2)}_{\pm}=\pm\frac{1}{2}\gamma\gamma_m\eps^{(2)}_{\pm}
\eea
Here $\gamma$ is a hermitean chirality matrix. The situation is different for an 
odd dimensional sphere which has only one class of Killing spinors (which will be 
denoted $\hat\eps$ in this appendix):
\bea\label{OddSphere}
\nabla_m{\hat\eps}_{\pm}=\pm\frac{i}{2}\gamma_m{\hat\eps}_{\pm}
\eea
Once we have a spinor on an odd--dimensional sphere $S^n$, it can be easily 
embedded in higher dimensions by using $n+p$ spit of gamma matrices:
\bea\label{AppSpinGamma}
\Gamma_m=\sigma\otimes \gamma_m,\quad \Gamma_\mu=\sigma_\mu\otimes 1,\quad 
\{\sigma,\sigma_\mu\}=0,\quad \mu=1,\dots, p
\eea
Unfortunately such simple decomposition does not work if $n$ is even and the goal 
of this appendix is to describe a construction of Killing spinors in that case. 
To do so we find it useful to study embeddings 
$S^n\rightarrow  S^{n+1}$ and extract some general lessons from this construction.

We begin with reviewing this construction for odd $n$. Writing the metric on 
$S^{n+1}$ as
\bea
d\Omega_{n+1}^2=d\theta^2+s_\theta^2 d\Omega_n^2,
\eea
we find the derivatives along $\Omega_n$ and $\theta$ directions:
\bea
\nabla_i\eps=\frac{i}{2}{\gamma}_i\eps-\frac{1}{2}c_\theta\sigma_\theta\sigma
{\tilde\gamma}_i\eps=
\frac{i}{2s_\theta}\gamma_i(\sigma-ic_\theta\sigma_\theta)\eps,\qquad
\nabla_\theta\eps=\d_\theta\eps
\eea
To reproduce the equation on the sphere we need to impose a projection:
\bea
0=(\sigma-ic_\theta\sigma_\theta+\la s_\theta\sigma\sigma_\theta)\eps=
e^{i\la\sigma\frac{\theta}{2}}(\sigma-i\sigma_\theta)e^{-i\la\sigma
\frac{\theta}{2}}\eps
\eea
With this projection we reproduce the "chiral" relation on $S^{n+1}$:
\bea
\nabla_i\eps=-\frac{i\la}{2}\gamma_i\Gamma\Gamma_\theta\eps,\quad 
\nabla_\theta \eps=\frac{i\la}{2}\Gamma\eps=
-\frac{i\la}{2}\Gamma_\theta\Gamma\Gamma_\theta\eps,
\quad
\eps=e^{i\la\Gamma\frac{\theta}{2}}\eps_0
\eea
Notice that by starting from ${\hat\eps}_+$ one can reproduce relations for both 
$\eps^{(2)}_+$ and 
$\eps^{(2)}_-$ Alternatively we could have started from ${\hat\eps}_-$ and produce
 both spinors on 
$S^{n+1}$. Such ambiguity is related to the fact that dimension of the spinor 
grows as we move from even to odd dimension. On the other hand, if we start from 
an even--dimensional sphere and add one more dimension, then the size of the 
spinor does not change and one needs to use both 
$\eps^{(2)}_+$ $\eps^{(2)}_-$ to construct a spinor in higher dimension. We will 
now describe the relevant procedure. 

We begin with "chiral" relation on even--dimensional sphere $S^{n}$ and write the 
derivative on $S^{n+1}$:
\bea
{\tilde\nabla}_i\eps_a=\frac{a}{2}{\tilde\gamma}_i\Gamma_\theta\eps_a:\qquad
\nabla_i\eps_a=\frac{a}{2s_\theta}\gamma_i\Gamma_\theta\eps_a-
\frac{c_\theta}{2s_\theta}\Gamma_\theta\gamma_i\eps_a=
\frac{1}{2s_\theta}\gamma_i\Gamma_\theta(a+c_\theta)\eps_a
\eea
Clearly this does not reduce to relation (\ref{OddSphere}), but we can define a new 
spinor
\bea
{\hat\eps}_+\equiv \eps_++\eps_-
\eea
then projection (\ref{EvenSphereChiral}) for ${\hat\eps}_+$ leads to the equation
\bea
(1+c_\theta-is_\theta\Gamma_\theta)\eps_+=-(-1+c_\theta-is_\theta\Gamma_\theta)
\eps_-:\qquad
(1+e^{-i\theta\Gamma_\theta})\eps_+=(1-e^{-i\theta\Gamma_\theta})\eps_-\nonumber
\eea
This equation and relation (\ref{OddSphere}) along $\theta$ direction can be solved 
simultaneously by expressing 
$\eps_+$ and $\eps_-$ in terms of $\theta$--independent spinor $\eps_0$:
\bea
\eps_+=i\sin\frac{\theta\Gamma_\theta}{2}\eps_0,\quad \eps_-=
\cos\frac{\theta\Gamma_\theta}{2}\eps_0,\quad
{\hat\eps}_+=\exp\left(\frac{i\theta\Gamma_\theta}{2}\right)\eps_0\equiv 
\eps_-+\Gamma_\theta{\tilde \eps}_+
\eea
This concludes the construction of a spinor on odd--dimensional sphere in terms of
 a "chiral" spinor on the even--dimensional one, but to learn a more general lesson
 about the spinors it is convenient to rewrite the above relation is a slightly 
different form. 

Let us go back to the equation for ${\eps}_\pm$ and combine them into a single 
relation:
\bea
{\tilde\nabla}_i(\eps_-+\Gamma_\theta{\tilde \eps}_+)=
\frac{1}{2}{\tilde\gamma}_i\Gamma_\theta(-\eps_-+\Gamma_\theta{\tilde \eps}_+)
\eea
This equation can be rewritten in terms of spinor $\eps_0$ and two matrices 
$\Gamma_\pm=\frac{1}{2}(1\pm\Gamma_\theta)$:
\bea
{\tilde\nabla}_i(e^{i\theta/2}\Gamma_++e^{-i\theta/2}\Gamma_-)\eps_0=
-\frac{1}{2}{\tilde\gamma}_i\Gamma_\theta
(e^{-i\theta/2}\Gamma_++e^{i\theta/2}\Gamma_-)\eps_0=
\frac{1}{2}{\tilde\gamma}_i
(-e^{-i\theta/2}\Gamma_++e^{i\theta/2}\Gamma_-)\eps_0\nonumber
\eea
In other words, we have two relations
\bea\label{AppSpin01}
{\tilde\nabla}_i\Gamma_+\eps_0=\frac{1}{2}{\tilde\gamma}_i\Gamma_-\eps_0,\quad
{\tilde\nabla}_i\Gamma_-\eps_0=-\frac{1}{2}{\tilde\gamma}_i\Gamma_+\eps_0
\eea
To summarize, we found that a spinor on an odd--dimensional sphere can be 
decomposed as 
\bea
{\hat\eps}_+=\frac{1}{2}(e^{i\theta/2}\Gamma_+\eps_0+e^{-i\theta/2}\Gamma_-\eps_0)
\eea
and spinor $\eps_0$ satisfies the equations (\ref{AppSpin01}). If we want to make 
the symmetries of $S^n$ explicit, it is convenient to choose a basis of gamma 
matrices which has a form (\ref{AppSpinGamma}). Strictly speaking, this cannot be 
done for even $n$ since the number of components of a spinor does not change as we
 go from even to odd dimension, however we can first double the size of the 
Killing spinor and then impose a projection. In the case of 
$S^n\rightarrow S^{n+1}$ lift we can choose a basis of gamma matrices 
\bea
\Gamma_i={\hat\sigma}_1\otimes {\tilde \Gamma}_i,\quad \Gamma_\theta=
{\hat\sigma}_3\otimes 1
\eea
and then require that Killing spinor satisfies a constraint which involves a 
chirality operator $\gamma$:
\bea 
{\hat\sigma}_3\otimes \gamma\cdot\eps=\eps
\eea
In particular, doubling the size of a spinor $\eps_0$ and imposing a constraint, 
one rewrites equations (\ref{OddSphere})
in terms of chirality matrix on $S^n$:
\bea\label{AppSpinMastEqn}
{\tilde\nabla}_i(1+\gamma)\eps_0=\frac{1}{2}{\tilde\gamma}_i(1-\gamma)\eps_0,\quad
{\tilde\nabla}_i(1-\gamma)\eps_0=-\frac{1}{2}{\tilde\gamma}_i(1+\gamma)\eps_0
\eea

While we used the lift $S^n\rightarrow S^{n+1}$ to motivate this relation, the 
result can be applied to a general embedding of even dimensional spheres into 
higher dimensional spaces. 
One first introduces a basis of gamma matrices:
\bea\label{EvenGamDecomp}
\Gamma_m=\sigma\otimes {\tilde\Gamma}_m,\quad \Gamma_\mu=\sigma_\mu\otimes 1,
\quad 
\{\sigma,\sigma_\mu\}=0,\quad \mu=1,\dots, p
\eea
The spinors are constrained by projection involving chirality operator $\gamma$ on
 the sphere:
\bea\label{EvenGamProj}
\sigma\otimes \gamma\cdot\eps=\eps
\eea
Finally, the equations along the sphere directions are given by 
(\ref{AppSpinMastEqn}). To use these equations, it is convenient to rewrite them 
as
\bea\label{SpinMaster}
{\tilde\nabla}_m\eps=-\frac{i}{2}({\sigma}{\tilde\gamma}_m) P\eps
\eea
and describe the properties of matrix $P$. First of all, since it contains a factor 
of $\sigma$, it anticommutes with all $\Gamma_\mu$. We will also need the 
commutation relation for $P$ and $\gamma$, and the simplest way to find it is to 
choose an explicit representation
\bea
\gamma=\left(\begin{array}{cc}1&0\\0&-1\end{array}\right)\otimes I
\eea
where $I$ is a unit matrix involving irrelevant components of the spinor. Then 
equations 
(\ref{AppSpinMastEqn}) imply that in this representation matrix $P$ has a form:
\bea
P=\sigma\otimes \left(\begin{array}{cc}0&-i\\i&0\end{array}\right)\otimes I
\eea
This shows that $P$ is hermitean matrix which anticommutes with $\gamma$. 

Let us now summarize the construction of Killing spinors on an even dimensional 
sphere. One defines a basis of gamma matrices (\ref{EvenGamDecomp}) and imposes a 
projection 
(\ref{EvenGamProj}) on a spinor.  Then the Killing spinor satisfies an equation 
(\ref{SpinMaster}) 
with hermitean matrix $P$ which anticommutes with $\gamma_\mu$ and with chirality 
operator $\gamma$. 

Finally, we make a brief comment about AdS space. All formulas for the spheres can
 be rewritten for this case using a simple analytic continuation, in particular 
in equation (\ref{SpinMaster}) a prefactor $-i/2$ should be replaced by $1/2$. 


\section{Perturbative solution at large distances.}
\renewcommand{\theequation}{C.\arabic{equation}}
\setcounter{equation}{0}

\label{AppPert}

In section \ref{SectPert} we outlined a procedure for constructing a solution as a
 perturbative series around $AdS_5\times S^5$. For all solutions which asymptote to
 $AdS_5\times S^5$ (i.e. for all functions $\Phi$ such that 
 $\d_y \Phi=0$ in a finite region of $y=0$ line) we expect this series to converge
 for large values of $x^2+y^2$, then the entire solution can be constructed as an 
analytic continuation of the series. Some intermediate steps were missing in 
section \ref{SectPert} and here we will fill the gaps. 

Let us consider a first order corrections to the fields, i.e. we take $m=1$ in 
equations (\ref{AsympExpans}). Then looking at definition of $\Psi_1$, 
$\Psi_2$, we find
\bea\label{Miracle}
d\Psi_2^{(1)}+*d\Psi_1^{(1)}=d(x\phi^{(1)})- *d(y\phi^{(1)})
\eea
Let us now expand the functions which enter equation (\ref{MastHarmonic}):
\bea
&&\arctan e^{G}=\arctan e^{G_0}+\frac{\eps g^{(1)}}{e^{G_0}+e^{-G_0}}=
\arctan e^{G_0}+\frac{\eps~ s~sh~g^{(1)}}{s^2+sh^2}\nonumber\\
&&F=\sqrt{c^2+\eps[ch^2~ h^{(1)}-sh^2~g^{(1)}+s^2~g^{(1)}-\frac{c^2}{2}\phi^{(1)}]}
\nonumber\\
&&\qquad=c\left[1-\frac{1}{4}\phi^{(1)}+
\frac{\eps}{2c^2}(h^{(1)}+(h^{(1)}-g^{(1)})sh^2+g^{(1)}s^2)\right]\equiv 
c(1+\eps f^{(1)})
\nonumber\\
&&e^A=ch\left[1-\eps\frac{\phi^{(1)}}{4}+\eps\frac{h^{(1)}}{2}\right]\nonumber\\
&&\frac{e^{A}-F}{e^A+F}=\frac{ch-c}{ch+c}\left[1+\eps\frac{2Fe^A}{e^{2A}-F^2}
(a^{(1)}-f^{(1)})\right]\nonumber\\ 
&&\left[\log\frac{e^{A}-F}{e^A+F}\right]^{(1)}=\frac{2Fe^A}{e^{2A}-F^2}(a^{(1)}-
f^{(1)})=\frac{x}{sh^2+s^2}\left[
h^{(1)}-\frac{h^{(1)}~ch^2}{c^2}+g^{(1)}\frac{sh^2-s^2}{c^2}
\right]\nonumber
\eea
Substituting this into the equations (\ref{MastHarmonic}), we arrive at the expressions:
\bea
&&\frac{g^{(1)}}{s^2+sh^2}-\phi^{(1)}=\frac{\d_y\Phi^{(1)}}{y},\nonumber\\
&&\frac{1}{2(sh^2+s^2)}\left[h^{(1)}-\frac{h^{(1)}ch^2}{c^2}+g^{(1)}
\frac{sh^2-s^2}{c^2}\right]+\phi^{(1)}=
\frac{\d_x\Phi^{(1)}}{x}
\eea
It is useful to rewrite the last equation in a different form:
\bea
\frac{\d_x\Phi^{(1)}}{x}&=&\frac{1}{2(sh^2+s^2)}\left[(g^{(1)}-h^{(1)})
\frac{sh^2+s^2}{c^2}-2g^{(1)}\frac{s^2}{c^2}\right]+\phi^{(1)}\nonumber\\
&=&-\frac{1}{2c^2}(h^{(1)}-g^{(1)}-2\phi^{(1)})+\frac{1}{2(sh^2+s^2)}
\left[-2g^{(1)}\frac{s^2}{c^2}\right]-
\phi^{(1)}\frac{s^2}{c^2}\nonumber\\
&=&-\frac{1}{2c^2}(h^{(1)}-g^{(1)}-2\phi^{(1)})-\frac{s^2}{c^2}
\frac{\d_y\Phi^{(1)}}{y}-\frac{2s^2}{c^2}\phi^{(1)}
\eea
At this point we can express everything in terms of $\Phi^{(1)}$ and $g^{(1)}$:
\bea\label{FrstOrdGH}
&&\phi^{(1)}=\frac{g^{(1)}}{s^2+sh^2}-\frac{\d_y\Phi^{(1)}}{y},\\
&&h^{(1)}-g^{(1)}-2\phi^{(1)}=-\left\{2s^2\frac{\d_y\Phi^{(1)}}{y}+
4s^2\phi^{(1)}+2c^2\frac{\d_x\Phi^{(1)}}{x}\right\}
\nonumber
\eea
To determine $g^{(1)}$ in terms of $\Phi^{(1)}$ we should use the equation
\bea
\d_y(h^{(1)}-g^{(1)}-2\phi^{(1)})=\frac{4y}{s^2+sh^2}\frac{g^{(1)}}{s^2+sh^2}
\eea
which is a counterpart of (\ref{YeqnMord}) for $m=1$. 
Evaluating the left--hand side of this relation:
\bea
\d_y(h^{(1)}-g^{(1)}-2\phi^{(1)})=-4\d_y \left(\frac{s^2g^{(1)}}{s^2+sh^2}\right)+
2\d_y\left[s^2\frac{\d_y\Phi^{(1)}}{y}-c^2\frac{\d_x\Phi^{(1)}}{x}\right],
\nonumber
\eea
we arrive at equation (\ref{FiEqnMord}) for $m=1$. It is clear that the same set of 
equations which we derived now for $m=1$ would hold for any $m$, the only 
difference would be in the source terms $\Phi^{(m)}$ and 
$\Psi_y^{(m)}$. 

We were able to derive an equation for the perturbation in a closed form in part 
due to a miraculous relation 
(\ref{Miracle}). We recall that in general to find the potentials $\Psi_1$ and 
$\Psi_2$ in terms of the metric 
and the dilaton one needs to solve differential equations, but in the leading 
order the relation $dx= *dy$ 
led to algebraic expression for the potentials. One can hope that similar 
algebraic relation persists to higher 
orders as well. In the remaining part of this appendix we will analyze this 
question for the second order in 
perturbation and we conclude that there is no relation of the type 
(\ref{Miracle}). Another purpose to present 
these calculations here is to provide expressions for the first order correction 
to the metric in a form which is 
more explicit than (\ref{MthOrderEqn}), (\ref{FiEqnMord}). 

Let us look at the second correction to (\ref{DefPsi}):
\bea\label{NotAMiracle}
d\Psi^{(2)}_2+*d\Psi^{(2)}_1&=&d(x\phi^{(2)})- *d(y\phi^{(2)})\nonumber\\
&&+\left(\frac{e^{A} F}{e^{2A}-e^{2B}}\right)^{(1)}
d\phi^{(1)}-y\left(\frac{e^{-\phi}}{e^{2A}-e^{2B}}\right)^{(1)}*d\phi^{(1)}
\eea
We want to see whether the second line of this relation admits a simple 
decomposition similar to the one in the first line. To answer this question we 
have to evaluate the second line and try to guess an appropriate decomposition. 
While we can construct a complete solution in the linear order starting from any 
harmonic 
function $\Phi$, such solutions look quite complicated due to presence of 
hyperbolic functions. One can first try 
to address the question in the "Poincare patch" of AdS space, i.e. in the region 
where hyperbolic functions can be replaced by the exponents. In this region the 
solution simplifies dramatically and we can write explicit 
expressions for all metric components. Notice that our perturbative expansion 
should work only at large values of $x^2+y^2$, i.e. precisely in the regime of 
validity of the "Poincare patch." However to compute the second order we would 
want to keep various modes of $\Phi$ (which go like $S_n e^{-n\rho+in\theta}$) at 
the same time, so one might think that approximation of hyperbolic functions is 
not a good idea. However if we are interested in contribution to the second line 
of (\ref{NotAMiracle}) which is proportional to $S_n S_m$, it is true that the 
leading contribution to this quantity is given by a Poincare patch result. 

As we mentioned before, to construct the first perturbative correction 
to $AdS_5\times S^5$ solution, 
one should start with harmonic  function $\Phi^{(1)}$, then find the corresponding
 fields $\phi^{(1)}$, $g^{(1)}$, 
$h^{(1)}$ using equations (\ref{FiEqnMord}), (\ref{MthOrderEqn}), 
then the fluxed can be recovered using 
(\ref{MastEqnF1})--(\ref{MastEqnF3}). Now we solve this system in 
the Poincare patch, i.e. we consider the region of large $\rho$ and replace 
hyperbolic functions by the 
exponents. Then it is convenient to start from a multipole expansion of the 
harmonic function $\Phi^{(1)}$:
\bea\label{AppEqn20}
\Phi^{(1)}=Q_0\rho+\sum_{n>0} Q_n(z) e^{-n\rho}:\quad (1-z^2)\d_z^2 Q_n-
z\d_z Q_n+n^2 Q_n=0,\quad 
z\equiv\cos\theta
\eea
Since we want this function to be suppressed compared to the $AdS_5\times S^5$ 
contribution (which naively goes like $e^\rho$), the index $n$ should be 
non--negative. However nontrivial  
$Q_0$ corresponds to changing the radius of the AdS space, so we will have to set 
it to zero. Also $Q_1$ corresponds to a dipole moment of the electrostatic problem, and since 
the total charge of the system is non--zero (it is related with radius of AdS), we can always make 
a shift in $x$ coordinate to set $Q_1(z)=0$. 
Notice that there one can also add a constant to $\Phi^{(1)}$, but it will not 
affect the solution. 

For large values of $\rho$, equation (\ref{FiEqnMord}) becomes:
\bea\label{AppEqn21}
\frac{c}{s}\d_y\left(\frac{s^3}{c}2e^{-2\rho} g^{(1)}\right)=
\frac{1}{4}\d_y\left\{\frac{s^2}{y}\d_y\Phi^{(1)}-\frac{c^2}{x}\d_x\Phi^{(1)}
\right\}\approx
\d_y\left\{e^{-2\rho}(y\d_y-x\d_x)\Phi^{(1)}\right\}
\eea
We observe that $Q_0$ does not source the correction to the metric $g^{(1)}$. Let 
us now introduce the mode expansion for the first order corrections:
\bea\label{ExpandModes}
g^{(1)}=\sum_{n>1} g_n(z)e^{-n\rho},\quad h^{(1)}=\sum_{n>1} h_n(z)e^{-n\rho},
\quad 
\phi^{(1)}=\sum_{n>1} P_n(z)e^{-(n+2)\rho}
\eea
We introduced a shift into the modes of dilaton due to the relation (\ref{MthOrderEqn}) 
which implies that in the Poincare 
patch there is a linear relation between $P_n$ and $Q_n$, $g_n$ once expansions 
(\ref{ExpandModes}) are 
defined. For our purposes it would be convenient to express all fields in terms 
of $P_n$ rather than $Q_n$ but 
$P_n$ can always be expressed in terms of $Q_n$ using (\ref{AppEqn21}). In 
particular, (\ref{AppEqn20}) would imply a second order differential equation for 
$P_n$, but to get this equation one needs a certain amount of a guesswork. Rather 
than taking this route, we use a different method which directly leads to an 
equation for 
$P_n$, and we will use (\ref{AppEqn21}) to relate $P_n$ and $Q_n$ in the end. 

In the approximation that we are taking, the equations for fluxes (\ref{MastEqnF1}), 
(\ref{MastEqnF2}) collapse to simple relations:
\bea
df^{(1)}_1=-\frac{e^{3\rho}}{4}[c~d\phi^{(1)}- s*d\phi^{(1)}],\quad
df^{(1)}_2= \frac{e^{3\rho}}{4}\left[c~d\phi^{(1)}-s*d\phi^{(1)}\right]
\eea
which imply that $f^{(1)}_1=f^{(1)}_2$. Substituting expansion of $\phi^{(1)}$ 
from (\ref{ExpandModes}), we find the expression for $f^{(1)}_1$:
\bea
df^{(1)}_1=-\frac{1}{4}\sum_n e^{(1-n)\rho}\left\{[-(n+2)cP_n+s^2P_n']d\rho+[-scP_n'-
(n+2)sP_n]d\theta\right\}
\eea
Integrability condition for this relation leads to a differential equation for 
$P_n$:
\bea\label{ODEforP}
(1-z^2)P_n''-5zP_n'-(4-n^2)P_n=0
\eea
and now we want to express everything in terms of this function. We begin with 
rewriting equation (\ref{FiEqnMord}) in terms of $\phi^{(1)}$ rather that $g^{(1)}$:
\bea
-4\d_y (s^2\phi^{(1)})-\frac{4y\phi^{(1)}}{s^2+sh^2}=
\frac{4\d_y\Phi^{(1)}}{s^2+sh^2}+
2\d_y\left[s^2\frac{\d_y\Phi^{(1)}}{y}+
c^2\frac{\d_x\Phi^{(1)}}{x}\right]
\eea
Going to large values of $\rho$, we find an equation which holds on the Poincare 
patch: 
\bea
-\frac{1}{2}\d_y (s^2\phi^{(1)})-e^{-\rho}s\phi^{(1)}=2e^{-2\rho}\d_y\Phi^{(1)}+
\d_y\left[e^{-2\rho}(y\d_y\Phi^{(1)}+x\d_x\Phi^{(1)})
\right]\nonumber
\eea
Expanding this equations in terms of modes, we can solve for $Q_n$ as a function 
of $P_n$:
\bea\label{QasP}
Q_n=\frac{1}{2(n-1)n^2}[z(1-z^2)P_n'+((2+n^2)(1-z^2)-3)P_n]
\eea
To arrive at this relation we used equation (\ref{ODEforP}). As a consistency check 
one can see that function $Q_n$ satisfies its equation (\ref{AppEqn20}) as long as 
$P_n$ satisfies (\ref{ODEforP})\footnote{The computation mentioned here can be easily 
performed in Mathematica.} One can also invert a relation (\ref{QasP}) to express 
$P_n$ through $Q_n$ and $Q'_n$, but the result is quite complicated, so we do not 
write it here. Finally we can use equations 
(\ref{FrstOrdGH}) to evaluate
\bea
g_n=h_n=\frac{n+1}{4n(n-1)}\left[-2z(1-z^2)P_n'-2(n+2)(1-z^2)P_n+
\frac{(n+2)(n+3)}{n+1}P_n\right]
\eea

Once we have a complete result for the solution in the first order, it can be 
used to compute the second 
correction to $\Psi_1$ and $\Psi_2$, in particular we want to check whether the 
simple algebraic relation 
similar to (\ref{Miracle}) persists at the second order.  The computations are 
straightforward but tedious, 
so we present only the essential steps. We begin with looking at the following 
expression and expand up to second order:
\bea\label{AppEqn456}
&&\frac{e^{-\phi/2}}{e^{2A}-e^{2B}}\left[e^{A+\phi/2} Fd\phi- y*d\phi\right]=
d(x\phi)-*d(y\phi)\nonumber\\
&&\qquad-\sum_{n>1}\frac{e^{(1-n)\rho}s}{8n}(2z(1-z^2)P_n'-
(n+2)(2z^2+1)P_n)*d\phi\\
&&+\sum_{n>1}
\frac{e^{(1-n)\rho}}{4}\left[\frac{(n+2)zP_n-(1-z^2)P_n'}{n-1}-
\frac{z(n+2)(1+2z^2)P_n-2z^2(1-z^2)P_n'}{2n}
\right]d\phi\nonumber
\eea
We now observe that there exists a set of harmonic functions $\zeta_n$:
\bea
\zeta_n\equiv \frac{1}{2}e^{(3-m)\rho}s\left[z(1-z^2)P_n'-((n+1)+(2-n)z^2)P_n\right]
\eea
and we also functions ${\tilde\zeta}_n$ which are dual to $\zeta_n$: 
\bea
d{\tilde\zeta}_n= *d\zeta_n:\qquad \d_\rho {\tilde\zeta}_n=
(1-n){\tilde\zeta}_n=\d_\theta\zeta_n
\eea
Plugging this into the (\ref{AppEqn456}), we find
\bea
&&\frac{e^{-\phi/2}}{e^{2A}-e^{2B}}\left[e^{A+\phi/2} Fd\phi- y*d\phi\right]=
d(x\phi)-*d(y\phi)\nonumber\\
&&\qquad+\sum e^{(1-n)\rho}\left[
\frac{{\tilde\zeta}_n}{2n}d\phi-\frac{\zeta_n}{2n}*d\phi
+\frac{1}{2}\left(\frac{z(3-4z^2)}{4}P_nd\phi-\frac{1-4z^2}{4}P_n*d\phi\right)\right]
\nonumber\\
&&=d\left(x\phi+\sum\frac{e^{(1-n)\rho}{\tilde\zeta}_n}{2n}\phi\right)-
*d\left(y\phi+\sum\frac{e^{(1-n)\rho}\zeta_n}{2n}\phi\right)\nonumber\\
&&\qquad+\sum\frac{e^{(1-n)\rho}}{8}P_n\left[
-(1-4z^2)s*d\phi+z(3-4z^2)d\phi\right]
\eea
Notice that due to the relations
\bea
\d_\theta(z(3-4z^2))=3s(-1+4z^2),\quad \d_\theta(s(1-4z^2))=3z(3-4z^2)
\eea
we have the duality
\bea
d(z(3-4z^2)e^{3\rho})=-*d[(-1+4z^2)e^{3\rho}]
\eea
this leads to equation
\bea
&&\frac{e^{-\phi/2}}{e^{2A}-e^{2B}}\left[e^{A+\phi/2} Fd\phi- y*d\phi\right]
=d\left(x\phi+\sum\frac{e^{(1-n)\rho}{\tilde\zeta}_n}{2n}\phi+
\frac{x\phi^2}{8}e^{2\rho}(3-4z^2)\right)\nonumber\\
&&\qquad-
*d\left(y\phi+\sum\frac{e^{(1-n)\rho}\zeta_n}{2n}\phi+\frac{y\phi^2}{8}
e^{2\rho}(1-4z^2)\right)\nonumber
\eea
At this point we already have the expressions for $\Psi_1^{(2)}$ and 
$\Psi^{(2)}_2$, but it turns out that using explicit form of the $\zeta$ and 
${\tilde\zeta}$ they can be rewritten in a simpler form
\bea\label{ModifPrepot}
\Psi_1&=&y\phi\left(\frac{1}{e^{2A}-e^{2B}}+
\sum\frac{e^{-n\rho}}{8}(4z^2-1)P_n\right)\nonumber\\
&=&
y\phi\left(\frac{1}{e^{2A}-e^{2B}}+\frac{e^{2\rho}}{8}(4z^2-1)\phi^{(1)}\right)
\nonumber\\
\Psi_2&=&\phi\left(\frac{Fe^A}{e^{2A}-e^{2B}}-
\sum\frac{xP_n}{8}e^{-n\rho}(3-4z^2)\right)\nonumber\\
&=&
\phi\left(\frac{Fe^A}{e^{2A}-e^{2B}}-
\frac{xe^{2\rho}}{8}\phi^{(1)}(3-4z^2)\right)
\eea
The brackets in the above equations contain two terms: the first term suggests a 
simple algebraic relation analogous to (\ref{Miracle}), but the second terms 
destroy such simple connection. If one could guess the general structure of such 
extra terms (and if such terms can be written down in terms of algebraic functions
 or derivatives of warp factors) one would be able to start from a harmonic 
function $\Phi$ and write a solution of 
the entire system. Unfortunately, equation (\ref{ModifPrepot}) seems to indicate 
that if such algebraic 
expressions for $\Psi_1$ and $\Psi_2$ exist, they would be quite complicated, so 
at present time we have to 
rely on perturbation theory to find the geometry.


\begin{thebibliography}{99}
%
\bibitem{mald}
J.~M.~Maldacena,
  ``The large N limit of superconformal field theories and supergravity,''
  Adv.\ Theor.\ Math.\ Phys.\  {\bf 2}, 231 (1998);
  Int.\ J.\ Theor.\ Phys.\  {\bf 38}, 1113 (1999), hep-th/9711200.
%
\bibitem{gkpw}
S.~S.~Gubser, I.~R.~Klebanov and A.~M.~Polyakov,
  ``Gauge theory correlators from non-critical string theory,''
  Phys.\ Lett.\ B {\bf 428}, 105 (1998), hep-th/9802109.\\
 E.~Witten,
  ``Anti-de Sitter space and holography,''
  Adv.\ Theor.\ Math.\ Phys.\  {\bf 2}, 253 (1998), hep-th/9802150.
%
\bibitem{AdSrev}
%
O.~Aharony, S.~S.~Gubser, J.~M.~Maldacena, H.~Ooguri and Y.~Oz,
  ``Large N field theories, string theory and gravity,''
  Phys.\ Rept.\  {\bf 323}, 183 (2000)
  [arXiv:hep-th/9905111].
  %
\bibitem{llm}
H.~Lin, O.~Lunin and J.~Maldacena,
  ``Bubbling AdS space and 1/2 BPS geometries,''
  JHEP {\bf 0410}, 025 (2004), hep-th/0409174.
%
\bibitem{jevicki}
S.~Corley, A.~Jevicki and S.~Ramgoolam,
  ``Exact correlators of giant gravitons from dual N = 4 SYM theory,''
  Adv.\ Theor.\ Math.\ Phys.\  {\bf 5}, 809 (2002), hep-th/0111222.
%
\bibitem{berenst}
D.~Berenstein,
  ``A toy model for the AdS/CFT correspondence,''
  JHEP {\bf 0407}, 018 (2004), hep-th/0403110.
%
\bibitem{giant}
J.~McGreevy, L.~Susskind and N.~Toumbas,
  ``Invasion of the giant gravitons from anti-de Sitter space,''
  JHEP {\bf 0006}, 008 (2000), hep-th/0003075;\\
 M.~T.~Grisaru, R.~C.~Myers and O.~Tafjord, ``SUSY and Goliath,''
  JHEP {\bf 0008}, 040 (2000), hep-th/0008015.
%
\bibitem{hashHirItz}
A.~Hashimoto, S.~Hirano and N.~Itzhaki,
 ``Large branes in AdS and their field theory dual,''
  JHEP {\bf 0008}, 051 (2000), hep-th/0008016.
%
\bibitem{mandal}
G.~Mandal,``Fermions from half-BPS supergravity,''
  JHEP {\bf 0508}, 052 (2005), hep-th/0502104.
%
\bibitem{maoz}
L.~Maoz and V.~S.~Rychkov,
  ``Geometry quantization from supergravity: The case of 'bubbling AdS',''
  JHEP {\bf 0508}, 096 (2005), hep-th/0508059.
%
\bibitem{reyYee}
S.~J.~Rey and J.~T.~Yee,
  ``Macroscopic strings as heavy quarks in large N gauge theory and  anti-de
  Sitter supergravity,''
  Eur.\ Phys.\ J.\ C {\bf 22}, 379 (2001), hep-th/9803001.
%
\bibitem{maldLoop}
J.~M.~Maldacena,
  ``Wilson loops in large N field theories,''
  Phys.\ Rev.\ Lett.\  {\bf 80}, 4859 (1998), hep-th/9803002.
%
\bibitem{CalMald}
C.~G.~.~Callan and J.~M.~Maldacena,
  ``Brane dynamics from the Born-Infeld action,''
  Nucl.\ Phys.\ B {\bf 513}, 198 (1998), hep-th/9708147.
%
\bibitem{yama}
S.~Yamaguchi,
  ``Bubbling geometries for half BPS Wilson lines,''
  arXiv:hep-th/0601089.
%
\bibitem{cardy}
J.~L.~Cardy,
  ``Conformal Invariance And Surface Critical Behavior,''
  Nucl.\ Phys.\ B {\bf 240}, 514 (1984).
%
\bibitem{kapustin}
A.~Kapustin,
 ``Wilson-'t Hooft operators in four-dimensional gauge theories and
 S-duality,'' arXiv:hep-th/0501015.
%
\bibitem{gomis}
S.~Yamaguchi,
  ``Wilson loops of anti-symmetric representation and D5-branes,''
  arXiv:hep-th/0603208;\\
 J.~Gomis and F.~Passerini, ``Holographic Wilson loops,''
  arXiv:hep-th/0604007.
%
\bibitem{fiol}
N.~Drukker and B.~Fiol,
  ``All-genus calculation of Wilson loops using D-branes,''
  JHEP {\bf 0502}, 010 (2005), hep-th/0501109.
%
\bibitem{douglas}
C.~Bachas, M.~R.~Douglas and C.~Schweigert,
  ``Flux stabilization of D-branes,''
  JHEP {\bf 0005}, 048 (2000), hep-th/0003037.
%
\bibitem{pawRey}
J.~Pawelczyk and S.~J.~Rey,
  ``Ramond-Ramond flux stabilization of D-branes,''
  Phys.\ Lett.\ B {\bf 493}, 395 (2000), hep-th/0007154.
%
\bibitem{kumar}
S.~A.~Hartnoll and S.~Prem Kumar,
  ``Multiply wound Polyakov loops at strong coupling,'' hep-th/0603190.
%
\bibitem{camino}
J.~M.~Camino, A.~Paredes and A.~V.~Ramallo,
  ``Stable wrapped branes,''
  JHEP {\bf 0105}, 011 (2001), hep-th/0104082.
%
\bibitem{vijay}
V.~Balasubramanian, M.~Berkooz, A.~Naqvi and M.~J.~Strassler,
  ``Giant gravitons in conformal field theory,''
  JHEP {\bf 0204}, 034 (2002), hep-th/0107119.
%
\bibitem{LinMald}
H.~Lin and J.~Maldacena,
  ``Fivebranes from gauge theory,''
  arXiv:hep-th/0509235.
%
\bibitem{skenTayl}
K.~Skenderis and M.~Taylor,
  ``Branes in AdS and pp-wave spacetimes,''
  JHEP {\bf 0206}, 025 (2002), hep-th/0204054.
%
\bibitem{mikhail}
A.~Mikhailov,
  ``Giant gravitons from holomorphic surfaces,''
  JHEP {\bf 0011}, 027 (2000), hep-th/0010206.
%
\bibitem{newBeren}
D.~Berenstein,
  ``Large N BPS states and emergent quantum gravity,''
  JHEP {\bf 0601}, 125 (2006), hep-th/0507203;\\
 D.~Berenstein, D.~H.~Correa and S.~E.~Vazquez,
  ``All loop BMN state energies from matrices,''
  JHEP {\bf 0602}, 048 (2006), hep-th/0509015.
%
\bibitem{dymar}
A.~Dymarsky, S.~Gubser, Z.~Guralnik and J.~Maldacena,
  arXiv:hep-th/0604058.
%
\bibitem{schwarz}
J.~H.~Schwarz,
  ``Covariant Field Equations Of Chiral N=2 D = 10 Supergravity,''
  Nucl.\ Phys.\ B {\bf 226}, 269 (1983);\\
 J.~H.~Schwarz and P.~C.~West,
  ``Symmetries And Transformations Of Chiral N=2 D = 10 Supergravity,''
  Phys.\ Lett.\ B {\bf 126}, 301 (1983);\\
 P.~S.~Howe and P.~C.~West,
  ``The Complete N=2, D = 10 Supergravity,''
  Nucl.\ Phys.\ B {\bf 238}, 181 (1984).
%
\bibitem{granaPolch}
M.~Grana and J.~Polchinski,
 ``Gauge / gravity duals with holomorphic dilaton,''
  Phys.\ Rev.\ D {\bf 65}, 126005 (2002), hep-th/0106014.
%
\bibitem{gaunt}
J.~P.~Gauntlett,
  ``Classifying supergravity solutions,''
  Fortsch.\ Phys.\  {\bf 53}, 468 (2005), hep-th/0501229;\\
J.~P.~Gauntlett, D.~Martelli, J.~Sparks and D.~Waldram,
  ``Supersymmetric AdS(5) solutions of type IIB supergravity,'' hep-th/0510125.
%
\bibitem{pope}
H.~Lu, C.~N.~Pope and J.~Rahmfeld,
  ``A construction of Killing spinors on $S^n$,''
  J.\ Math.\ Phys.\  {\bf 40}, 4518 (1999), hep-th/9805151.
%


\end{thebibliography}
\end{document}